\documentclass[reprint,amsmath,amssymb,aps,pra,superscriptaddress
]{revtex4-2}

\usepackage{graphicx}
\usepackage{dcolumn}
\usepackage{bm}
\usepackage{hyperref}
\usepackage{color}
\usepackage{xcolor}

\usepackage{physics}
\DeclareMathOperator{\sinc}{sinc}

\begin{document}

\preprint{APS/123-QED}

\title{Efficient optical cat state generation using squeezed few-photon superposition states}

\author{Haoyuan Luo}
\email{haoyuan.luo@sydney.edu.au}
\affiliation{
Centre for Engineered Quantum Systems, School of Physics, The University of Sydney, NSW 2006, Australia
}
\affiliation{Sydney Quantum Academy, Sydney, NSW, Australia}
\affiliation{%
 Institute for Photonics and Optical Sciences (IPOS), School of Physics, The University of Sydney, NSW 2006, Australia
}%
\author{Sahand Mahmoodian}%
\email{sahand.mahmoodian@sydney.edu.au}
\affiliation{
Centre for Engineered Quantum Systems, School of Physics, The University of Sydney, NSW 2006, Australia
}
\affiliation{%
 Institute for Photonics and Optical Sciences (IPOS), School of Physics, The University of Sydney, NSW 2006, Australia
}%

\begin{abstract}
Optical Schr\"{o}dinger cat states are non-Gaussian states with applications in quantum technologies, such as for building error-correcting states in quantum computing. Nevertheless, the efficient generation of high-fidelity optical Schr\"{o}dinger cat states is an outstanding problem in quantum optics. Here, we propose using squeezed superpositions of zero and two photons, $|\theta\rangle = \cos{(\theta/2)}|0\rangle + \sin{(\theta/2)}|2\rangle$, as ingredients for protocols to efficiently generate high-fidelity cat states. We present a protocol using linear optics with success probability $P\gtrsim 50\%$ that can generate cat states of size $|\alpha|^2=5$ with fidelity $F>0.99$. The protocol relies only on detecting single photons and is remarkably tolerant of loss, with $2\%$ detection loss still achieving $F>0.98$ for cats with $|\alpha|^2=5$. We also show that squeezed $\theta$ states are ideal candidates for nonlinear photon subtraction using a two-level system with near deterministic success probability and fidelity $F>0.98$ for cat states of size $|\alpha|^2=5$. Schemes for generating $\theta$ states using quantum emitters are also presented. Our protocols can be implemented with current state-of-the-art quantum optics experiments.
\end{abstract}

\maketitle


\section{Introduction}
Efficiently generating high-fidelity, large-photon-number non-Gaussian states of propagating photons remains an outstanding problem in quantum optics. The optical Sch\"{o}dinger cat state, a superposition of opposite amplitude coherent states $\ket{\text{cat}_{\alpha,\pm}}\propto\ket{\alpha}\pm\ket{-\alpha}$, is a particularly sought after non-Gaussian state as it forms a building block to generate other more complex non-Gaussian states such as the Gottesman-Kitaev-Preskill (GKP) state \cite{GKP}. The GKP state has been proposed as a candidate for encoding qubits in fault-tolerant quantum computing and can be generated by `breeding' squeezed cat states \cite{Terhal2018PRA,furusawa_gkp}. Large-amplitude cat states are therefore considered to be a critical resource state for photonic quantum computing \cite{Jeong2002PRA,tim_ralph_quantum_computation,Bergmann2016PRA,Li2017PRL,Xu2023npj,Schlegel2022PRA,Le2018PRL,Grimsmo2020PRX,Chamberland2022PRXQ,Gravina2023PRXQ,Lee2013PRA,Lee2024PRXQ,tim_ralph_cat,Albert2016PRL}.

\begin{figure}[t]
    \centering
    \includegraphics[width=\columnwidth]{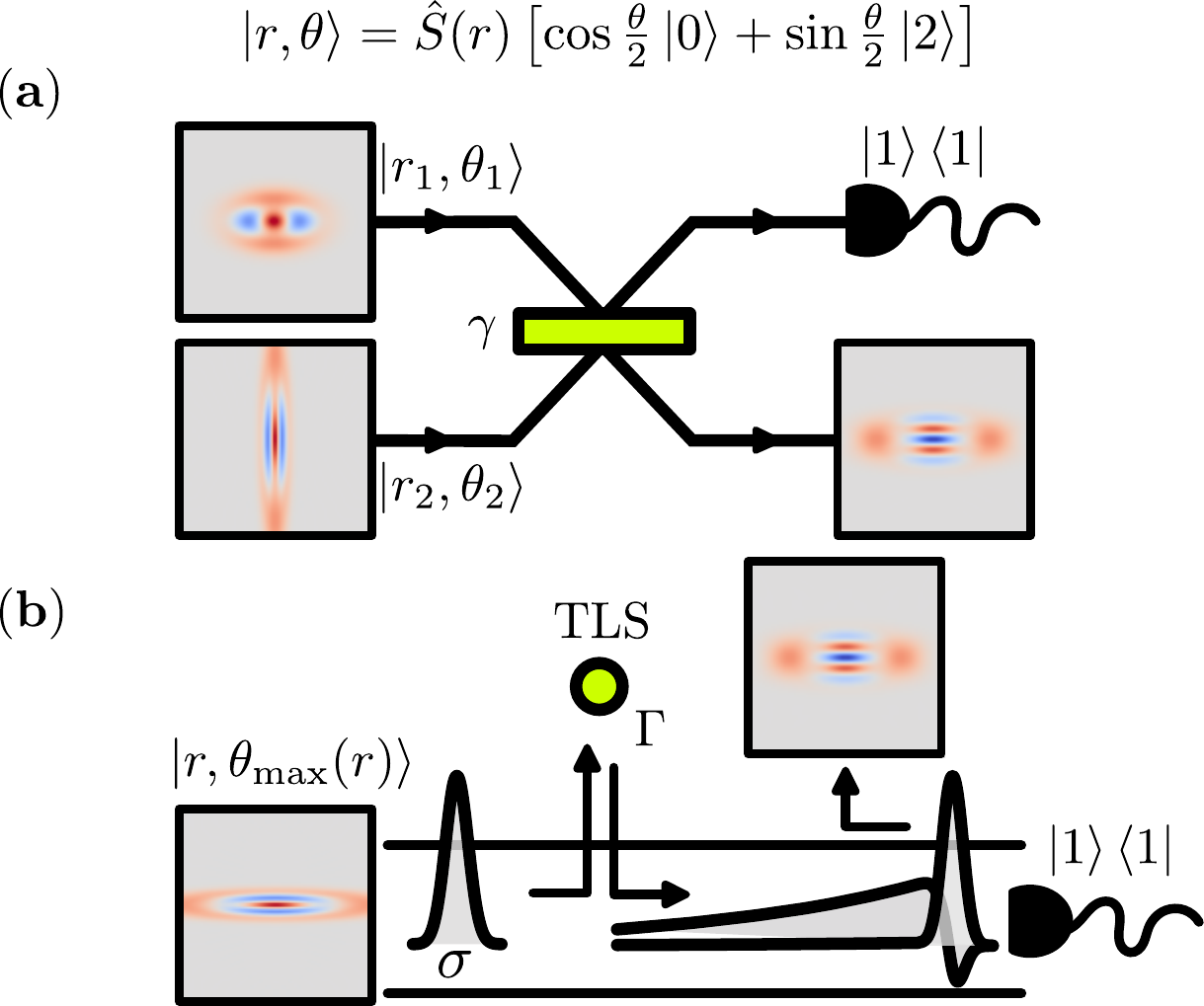}
    \caption{\label{fig:schematic}Schematics of linear (a) and nonlinear (b) single-photon subtraction schemes for cat state generation.  (a) Two resources states, parametrized by their squeezing strengths $r_i$ and angles $\theta_i$, are used as inputs into a beamsplitter with transmission and reflection coefficients $\cos\gamma$ and $i\sin\gamma$, respectively. Conditioned on a single-photon detection in the top output channel, a cat state is prepared in the bottom channel. (b) A resource state with $\theta_{\text{max}}(r)$ chosen such that the mean photon number is maximised under squeezing $r$ is prepared in a Gaussian wave packet with width $\sigma$ and scatters off a TLS that is chirally coupled to the waveguide with coupling strength $\Gamma$. After scattering, the most dominant mode is extracted with the use of a quantum pulse gate. Detection of a single photon in the leftover modes heralds a cat state in the dominant mode. The input and output states are depicted with use of their Wigner functions. After unsqueezing (not shown), the output states here are approximately cat states of size $\abs{\alpha}^2=4$.}
    \label{fig:schematic}
\end{figure}

Since cat states are non-Gaussian bosonic states, their generation either requires a non-Gaussian input state \cite{Sychev2017Npho,cat_from_photonnumberstates,amplify_large_cat,amplify_large_cat_2} or requires a Gaussian input state to undergo non-Gaussian evolution \cite{Weedbrook2012RMP,Chabaud2020PRL}. In circuit quantum electrodynamics platforms, large-amplitude cat states can be generated by means of nonlinear interactions present in the strong-dispersive regime \cite{Vlastakis2013SCIENCE}; however, such techniques cannot be implemented in optical systems due to weaker relative coupling strengths \cite{Schuster2007Nature,Najer2019Nature}. Therefore, common approaches rely on using photon-number detection as a nonlinearity to perform photon subtraction on a squeezed vacuum \cite{first_cat, kitten_experiment, molmer_squeeze_cat, cat_photon_subtraction_1, cat_photon_subtraction_2, furusawa}. Photon subtraction protocols usually require the use of linear optics to interfere photons \cite{photon_subtraction_experiment,photon_subtraction_experiment2}; however, the success probability of generating cat states of large amplitudes is hindered by the conditional nature of these protocols. To tackle this high degree of nondeterminism, alternative protocols involving multiple stages of photon subtraction were proposed \cite{cluster_state_cat,tim_ralph_cat}, resulting in near-deterministic cat state generation. However, the output fidelity is highly sensitive to detector efficiencies \cite{tim_ralph_cat}, and the cat state sizes produced in Ref.~\cite{cluster_state_cat} are probabilistic. 

In this paper, we propose interfacing few-photon superposition states, which can potentially be generated rapidly by quantum emitters, with squeezing to efficiently generate high-fidelity cat states. Single-photon sources have matured greatly in recent years, and emitters such as quantum dots and atoms coupled to cavities can efficiently produce high-quality single-photon states \cite{Tomm2021NNANO, Thomas2022Nature}. We propose using quantum emitters to deterministically generate superpositions of zero and two photons, $\ket{\theta} = \cos{(\theta/2)} \ket{0} + \sin{(\theta/2)} \ket{2}$. These resource states are already non-Gaussian, but have limited photon number and become poor approximations for larger cat states \cite{Takeoka2008PRA,Marek2008PRA}. To use these to produce cat states, we consider squeezing the state with up to $10$~dB of squeezing and performing a photon subtraction protocol. Here we present two photon subtraction protocols. The first protocol (Fig.~\hyperref[fig:schematic]{1(a)}) uses two squeezed $\theta$ states and implements photon subtraction using linear optics. The second protocol (Fig.~\hyperref[fig:schematic]{1(b)}) requires only a single squeezed $\theta$ state, and photon subtraction is implemented with a two-level system (TLS). We show that the use of a $\theta$ state as an input, as opposed to a Fock state, enables us to have a simultaneously high success probability and high fidelity, something that squeezed two-photon Fock state inputs cannot produce. Although our protocols require in-line squeezing, they otherwise provide a balance of minimising experimental resources while having high success probability $P_{\rm suc}\gtrsim50\%$, and produce cat states with $|\alpha|^2 >4$ with fidelity $F>0.99$.

\section{$\theta$ states}
Since cat states are supported on either the even-number basis or odd-number basis, for any photon subtraction protocol, one must fix the parity of the supported number basis of the resource state. The squeezing operator, defined as $\hat{S}(r)=e^{-\frac{r}{2}(\hat{a}^2-\hat{a}^{\dagger 2})}$ with squeezing strength $r\in\mathbb{R}$, preserves the parity of any particular number state. This reasoning underlies the use of a squeezed vacuum and photon subtraction to generate cat states. Since we propose using quantum emitters to deterministically generate resource states, we seek to take advantage of a squeezed superposition of a vacuum and a two-photon Fock state---namely,
\begin{align}
\label{theta state}
   \ket{r,\theta}=\hat{S}(r)\left[\cos{\left(\frac{\theta}{2}\right)} \ket{0}+\sin{\left(\frac{\theta}{2}\right)}\ket{2} \right],
\end{align}
where $\theta\in[0,2\pi)$. These types of state were first considered in Refs.~\cite{Takeoka2008PRA,Marek2008PRA}. The suitability of $\ket{r,\theta}$ for approximating cat states can be understood in terms of the stellar rank or the core-state expansion \cite{Chabaud2020PRL, Menzies2009PRA, Tzitrin2020PRA, Walschaers2021PRXQ}. Gaussian states have zero stellar rank and Gaussian processes do not change the stellar rank of a state. On the other hand, non-Gaussian states, such as cat states, have infinite stellar rank. Nevertheless, such non-Gaussian states can be well-approximated by squeezing a finite superposition of Fock states, which itself has a finite stellar rank. The quality of this approximation improves as the Fock state superposition is truncated at higher values, i.e., by squeezing a state with a larger stellar rank. The state in Eq.~\eqref{theta state} can therefore be made to better approximate larger Schr\"{o}dinger cat states after undergoing a non-Gaussian process such as photon subtraction.

\section{Linear photon subtraction}
\label{sec: linear}
\begin{figure}[t]
    \centering
    \includegraphics[width=\columnwidth]{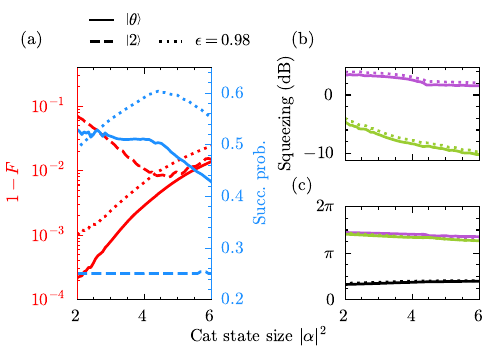}
    \caption{Performance of and parameters for the linear photon subtraction protocol. (a) Optimised infidelity (red lines) and success probability (blue lines) for generating cat states of size $\abs{\alpha}^2\in[2,6]$ with the squeezed $\theta$ states (solid lines) and squeezed two-photon states (dashed lines) as inputs. (b) Squeezing strengths $r_1$ (purple lines) and $r_2$ (green lines). Note the signs of the squeezing strengths correspond to squeezing in perpendicular directions. (c) Angles $\theta_1$ (purple lines) and $\theta_2$ (green lines) and the beamsplitter parameter $\gamma$ (black lines). In (a-c) we include the reoptimised results and parameters of the $\theta$ states with detector efficiency of $\epsilon=0.98$ (dotted lines).}
    \label{fig:fid_prob}
\end{figure}

We propose using squeezed $\theta$ states as inputs for generalized linear photon subtraction (see Fig.~\hyperref[fig:schematic]{1(a)}). In this process, two different squeezed $\theta$ states are combined with the use of a beamsplitter and photon subtraction is heralded by the detection of a single photon in one of the output channels. For optimized parameters, this process yields a cat state in the other output channel. This process produces a squeezed cat state that can be unsqueezed if necessary. A key distinction between `breeding' cat states and our protocol is the former relies on a Gaussian homodyne measurement \cite{Sychev2017Npho}, while our protocol uses single-photon detection which increases the output stellar rank \cite{Chabaud2020PRL}, potentially producing larger states. The parameters for this setup are as follows: the squeezing strengths $r_1, r_2$, the angles for the input states $\theta_1, \theta_2$, the beamsplitter transmissivity represented by the angle $\gamma$ and the unsqueezing strength on the output state. We optimized these parameters to yield a large success probability and fidelity, defined as $F=\abs{\braket{{\rm cat}_{\alpha,-}}{\rm out}}^2$, against a target cat state of size $\abs{\alpha}^2\in[2,6]$. The infidelities and success probabilities are presented in Fig.~\hyperref[fig:fid_prob]{2(a)}.

The performance of the $\theta$ states can be interpreted by the stellar ranks of the output states. The output state continuous variable wavefunction is expressed as $\psi_{\rm out}(x)\propto (b_1x+b_3x^3+b_5x^5)e^{-cx^2}$ (see Appendix~\ref{appendix: linear photon subtraction} for the fidelity, success probability, and stellar rank), and the stellar rank is 5. In the cases when the squeezing are equal in magnitude but opposite in sign, or one of the input states is in a vacuum, $b_5=0$ and the stellar rank is 3. When both conditions are simultaneously satisfied, $b_3=b_5=0$ and the stellar rank is 1. Note that if both inputs are a vacuum ($\theta_{1,2}=0$), but under different squeezing strengths, $b_3=b_5=0$, and our setup reduces to the generalized photon subtraction in Ref.~\cite{furusawa} and the stellar rank is 1. At first sight, it appears that using the $\theta$ state offers no advantage over using two-photon inputs as they have the same stellar rank. However, when the input states are fixed to two-photon Fock states, there is less flexibility for tuning the output wavefunction for higher success probabilities compared with $\theta$ states as inputs (see Appendix~\ref{appendix:theta state}). In Fig.~\hyperref[fig:fid_prob]{2(a)}, we compare the performance of using the $\theta$ states with that of using the two-photon Fock states as inputs. When $P_{\rm suc}\geq0.25$, the two-photon Fock states can prepare only poor-quality cat states. This calculation was performed with the analytic expressions for the fidelity and success probability (see Appendix~\ref{appendix: linear photon subtraction}). On the other hand, when the $\theta$ states are used, $F \geq 0.999$ for a cat state of size $\abs{\alpha}^2\leq 3.14$, and $F \geq 0.99$ for a cat state of size $\abs{\alpha}^2 \leq 5.40$. The success probability ranges from 0.529 to 0.427, which is almost twice as large as when the squeezed two-photon input states are used. We also note that for the same parameters, detection of two photons also produces a high-fidelity even-cat state, $\ket{{\rm cat}_{\alpha,+}}$ (see Appendix~\ref{appendix: linear photon subtraction}). Ultimately, the finite stellar rank of the $\theta$ states limits the size of the generated cat states, as presented by Fig.~\hyperref[fig:fid_prob]{2(a)}, and the fidelity drops below 0.99 for cat states with $|\alpha|^2>6$. Therefore, to generate cat states with even larger amplitudes, one should feed smaller cat states into enlargement protocols \cite{Sychev2017Npho,amplify_large_cat,amplify_large_cat_2} (see Appendix~\ref{appendix: linear photon subtraction}). Since these protocols are also probabilistic, the increase of the success probability of generating smaller cat states by a factor of 2 over two-photon inputs can lead to much higher compounded total success probability.

In state-of-the-art experiments, the main sources of imperfections are photon loss and imperfect detectors. We model both by inserting a linear loss element before the photon detector. The success probability and infidelity after reoptimisation of the parameters with the presence of loss are shown in Fig.~\hyperref[fig:fid_prob]{2(a-c)}. The protocol performs remarkably well with $2\%$ loss, with infidelity of $1\%$ or less for cat states with $|\alpha|^2 \leq 4$. We note that this is a significant loss tolerance compared with that of other protocols \cite{tim_ralph_cat}, with the robustness regarding loss being due to the protocol being heralded by detection of only a single photon.

\section{Nonlinear photon subtraction}
Although our linear optics photon subtraction protocol has high fidelity and success probability, it uses two squeezed $\theta$ states as inputs. We now consider a nonlinear photon subtraction protocol of a single squeezed $\theta$ state using a TLS chirally coupled to a waveguide. This type of photon subtraction protocol was recently proposed with squeezed vacuum inputs \cite{molmer_photon_subtraction}. We model the TLS-waveguide system under the Born-Markov approximation and the rotating-wave approximation together with renormalisation of the energy of the photon field to the resonance frequency of the TLS. The Hamiltonian for this system is ($\hbar=v_g=1$)
\begin{align}
\label{chiral atom hamiltonian}
    \hat{H}=&-i\int_{\mathbb{R}} dx \hat{a}^\dagger(x) \partial_x \hat{a}(x)+\sqrt{\Gamma}[\hat{a}^\dagger(0)\hat{\sigma}_-+\hat{\sigma}_+\hat{a}(0)],
\end{align}
where $\hat{a}^\dagger(x)$ is the photon creation operator at position $x$, $\hat{\sigma}_+$ and $\hat{\sigma
}_-$ are the Pauli operators of the TLS, and $\Gamma$ is the coupling strength \cite{quantum_chiral_optics,bound_states_dynamics}. To model the multi-mode dynamics of the photon field, we adapt the discretised input-output theory in the language of matrix product states (MPSs) \cite{time_bin_mps1,time_bin_mps2,zoller_mps}. In the narrow discretisation limit, the solution to the quantum stochastic Schr\"{o}dinger equation is well-approximated by a product of the first-order-trotter-decomposed time-evolution operators (see Appendix~\ref{appendix: nonlinear photon subtraction}).

In the nonlinear photon subtraction protocol, the TLS scatters an input wave packet into a product state composed of an output state of interest and a single-photon state in an orthogonal temporal mode, i.e., $\ket{\rm in}\rightarrow \ket{\rm out}\ket{1}$. For an $n$-photon Fock state this is almost deterministic \cite{split_two_photon,molmer_photon_subtraction,bound_states_dynamics}. However, it was shown that photon subtraction from a squeezed vacuum was not deterministic \cite{molmer_photon_subtraction}. This is because the squeezed vacuum pulses with a temporally narrow width required for ideal photon subtraction cannot deterministically scatter photons into an orthogonal mode. Our nonlinear single-photon subtraction protocol is shown in Fig.~\hyperref[fig:schematic]{1(b)}. A single-mode resource state $\ket{r,\theta_{\text{max}}}$ scatters into a multimode state, under the appropriate wave packet width and squeezing strength, and a cat state in the dominant mode is heralded by detection of a photon from the non-dominate modes. For the largest success probability we consider input states with the largest photon number. That is, we solve for $\theta_{\text{max}}(r)=\pi-\arctan(t_r/\sqrt{2})$, such that under squeezing, the mean photon number $\expval{\hat{n}}=c_r[\sqrt{2/(t_r^2+2)}+3/2]+s_rt_r/\sqrt{2(t_r^2+2)}-1/2$ of the resource state is maximized; here $c_r=\cosh(2r),s_r=\sinh(2r)$ and $t_r=\tanh(2r)$. 
We emphasize that this does not occur at $\theta=\pi$, highlighting the importance of generating $\theta$ states. Moreover, we numerically recover $\theta_{\rm max}$ when $\theta$ is included in one of the parameters in the optimization process (see Appendix~\ref{appendix: nonlinear photon subtraction}). Since the total number of excitations in the waveguide is conserved in the scattering process, $\theta_{\rm max}(r)$ also maximizes the mean photon number in the dominant mode, leading to the largest generated cat state.

To implement nonlinear photon subtraction we launch the input state $\ket{r,\theta_{\text{max}}}$ into a Gaussian wave packet and compute the scattering process using the MPS formalism (see Appendix~\ref{appendix: nonlinear photon subtraction}). We find the dominant mode of the output state by computing and diagonalizing the first-order coherence $G^{(1)}(t,t')=\expval{\hat{a}^\dagger(t)\hat{a}(t')}=\sum_i\Bar{n}_iv^*_i(t)v_i(t')$, where $v_i(t)$ are a set of orthonormal modes and $\Bar{n}_i$ are the corresponding mean photon numbers in each mode \cite{input_output_quantum_pulses}. This allows us to extract the dominant mode out of the waveguide with a virtual cavity \cite{input_output_quantum_pulses}. In an experiment, this is performed with a quantum pulse gate (QPG) \cite{Serino2023PRX}. Furthermore, we project the waveguide onto the single-photon sector by tracing over all possible single-photon detection events. The protocol thus assumes photon-number-resolving detection but does not require any timing resolution. Lastly, the state in the virtual cavity is the output state of the protocol.
Ideally, the resource state scatters into a two-mode product state, where the second-most dominant mode contains a single photon, thus making the protocol deterministic. However, with the presence of a vacuum and higher-photon-number subtraction events, ultimately, the protocol becomes nearly deterministic.

\begin{figure}[t]
    \centering
    \includegraphics[width=\columnwidth]{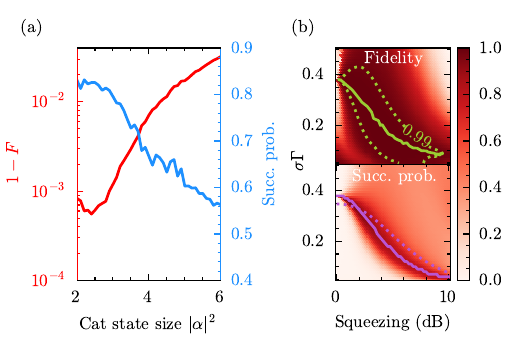}
    \caption{Fidelity and success probability for the nonlinear photon subtraction scheme. (a) Best infidelity (red line) and success probability (blue line) for generating a cat state of size $\abs{\alpha}^2$. (b) Fidelity (top) and success probability (bottom) across various input widths $\sigma\Gamma$ and squeezing strengths $r$. The dashed green contour encloses regions with fidelity $F>0.99$ and the solid green (purple) line highlights the best fidelity (success probability). The dotted purple line shows the values for inversion of the TLS with a $\pi$-pulse model.}
    \label{fig:nonlinear}
\end{figure}

We numerically sample across various wave packet widths and squeezing strengths and demonstrate regions of high success probability ($P_{\rm suc}>0.8$) of single-photon subtraction, see Fig.~\hyperref[fig:nonlinear]{3(b)}. The scattered photon field is approximately occupied by two modes (a dominant mode captured by the virtual cavity and an ancillary mode in the field). Consequently, tracing over all single-photon detection events in the field maintains high purity of the output state (see Appendix~\ref{appendix: nonlinear photon subtraction}). In Fig.~\hyperref[fig:nonlinear]{3(b)}, we show the fidelity of the virtual cavity (dominant mode), defined as $F=\bra{{\rm cat}_{\alpha,-}}\hat{\rho}_{\rm out}\ket{{\rm cat}_{\alpha,-}}$, against an optimally chosen cat state. Moreover, region of high fidelity ($F>0.99$) is enclosed by the dashed green contour and the highest fidelity is highlighted by the green line. Remarkably, the line of the highest success probability (purple) overlaps the line of the highest fidelity for squeezing strength up to 7~dB, corresponding to a generated cat state of size $\abs{\alpha}^2$ up to approximately 3. To generate larger cat states, the success probability is sacrificed to achieve optimal fidelity. 

The interplay between the input Gaussian width and the squeezing strengths can be understood by a simple $\pi$-pulse model, which models the excited state population of the TLS as though it is driven by a coherent field \cite{molmer_photon_subtraction}. It can predict the optimal pulse width ($\sigma_{\rm opt}\sim1/n$) for single-photon subtraction for high-photon-number Fock states \cite{molmer_photon_subtraction} by finding squeezing and $\sigma$ values such that $2 \int dt \sqrt{\Gamma \langle \hat{n}(t) \rangle}=\pi$; the regions of high success probability are near where the TLS is inverted by the input field. Beyond the $\pi$-pulse model, the strategy for achieving high success probabilities under small squeezing strengths is to minimize the amplitude of the vacuum, and indeed $\theta_{\rm max}$ approximates that up to the second order in $r$ (see Appendix~\ref{appendix: nonlinear photon subtraction}). On the other hand, under large squeezing strengths, the best strategy is to occupy high-photon-number Fock states, where the optimal width 
($\abs{\partial_n \sigma_{\rm opt}}\sim1/n^2$) is slowly varying compare with that of the few-photon Fock states \cite{molmer_photon_subtraction}. For a fixed squeezing strength, $\theta_{\rm max}$ precisely maximises the mean photon number and therefore  minimizes the change of the optimal width with respect to the $n$-photon Fock state. 

In Fig.~\hyperref[fig:nonlinear]{3(a)}, we showcase the best infidelity and the corresponding success probability for generating cat states of size $\abs{\alpha}^2\in[2,6]$. Again, the characteristic trade-off between fidelity and success probability is present; nonetheless, our protocol managed $F\geq0.99$ and $P_{\rm suc}\geq0.664$ for cat state size $\abs{\alpha}^2\leq4.3$. For larger cat states, the fidelity is ultimately limited by the stellar rank of the input state, which, after squeezing and photon subtraction, allows high-fidelity approximations only for cats states with $\abs{\alpha}^2\lesssim 4$. In contrast to the linear protocol introduced in earlier Sec.\ref{sec: linear}, the nonlinear protocol has higher success probabilities and uses only a single $\theta$ state.

\section{Preparation of the $\theta$ state}
We first consider a linear optics implementation that probabilistically heralds the generation of a $\theta$ state using single-photon states as inputs. Consider a beamsplitter setup similar to the one in Fig.~\hyperref[fig:schematic]{1(a)}, but with states $\ket{\phi_i} = \cos{(\phi_i/2)} \ket{0} + \sin{(\phi_i/2)} \ket{1}$ input in the top $i=1$ and bottom $i=2$. Such superposition states can easily be generated from any TLS photon source by preparing it in a superposition of excited and ground states. After these two states mix at a 50:50 beamsplitter the generation of a $\ket{\theta}$ state is heralded by the absence of photon clicks in the output detector. With state-of-the-art circuits, beamsplitters, and detectors, photon loss---which spoils the fidelity---can be made very small \cite{PsiQuantum2024Nature,Xanadu2025Nat}. For the correct state to be produced, we require $\phi_2 = - \phi_1$, which produces $\ket{\textrm{out}} = [\cos^2{(\phi_1/2)}|0 \rangle - \sin^2{(\phi_1/2)}\ket{2}/\sqrt{2}]/\sqrt{P_{\rm suc}}$ with success probability $P_{\rm suc}(\phi_1) = \cos^4{(\phi_1/2)} + \sin^4{(\phi_1/2)}/2$, where $1/3 \leq  P_{\rm suc} \leq 1$ depending on the desired angle of the $\theta$ state. To obtain a desired angle $\theta$, we find the correct value of $\phi_1$ by solving the equation $\cos{(\theta/2)} = \cos^2{(\phi_1/2)}/\sqrt{P_{\rm suc}(\phi_1)}$. Finally, we note the minus sign in $\ket{\textrm{out}}$, which is different from the $\theta$ state definition. This can be compensated for by adding a $\pi/2$ phase element to the optical path when $0\leq \theta \leq \pi$ is required.

We now consider deterministic generation using two three-level atoms coupled to an optical cavity with equivalent couplings. The three-level atoms each have a ground state $\ket{g}$ optically coupled to an excited state $\ket{e}$, and a shelving state $\ket{s}$. We can use this system to generate $\ket{\theta}$ by first preparing the two atoms in the state $(\cos{(\theta/2)}\ket{g} + \sin{(\theta/2)} \ket{s})\ket{g}$. We then perform a controlled-NOT gate within the $s$, $g$ subspace where $\ket{g}$ ($\ket{s}$) encodes the 0 (1) state. The first atom represented by the first ket acts as the control qubit and the second represents the target. Here, atomic systems with Rydberg transitions are ideal for implementing such a gate with very high fidelity \cite{Levine2019PRL}. This prepares the state $(\cos{(\theta/2)}\ket{gg} + \sin{(\theta/2)} \ket{ss})$. By performing an optical $\pi$ pulse on the $s$-$e$ transition, the population is moved coherently from $s$ to $e$, and the atoms then decay along the $e$-$g$ transition. The atom-photon system is left in the state $\ket{\bar{\theta}}\ket{gg} = (\cos{(\theta/2)}\ket{0} + \sin{(\theta/2)}\ket{\bar{2}} \ket{gg}$. We note that the two-photon state $\ket{\bar{2}}$ is not a Fock state, but is rather a superradiant state (see Appendix~\ref{appendix: theta state generation}) with two-photon wavefunction (for $x_1 \geq x_2$),
\begin{align}
    \psi(x_1, x_2) = \sqrt{2}\Gamma \theta(-x_1)\theta(-x_2) e^{\Gamma x_2/2},
\end{align}
where $\theta(x)$ is the Heaviside step function. Since our protocols require single-mode states, we calculate the fidelity of the state $\ket{\bar{\theta}}$ by finding the dominant mode of $\psi(x_1,x_2)$ and computing $F = |\braket{\theta}{\bar{\theta}}|^2$. We find that the infidelity $1 - F \sim 2 \sin^2{(\theta/2)}0.046$ with $0.046 = 1- \braket{2}{\bar{2}}$ corresponding to a minimal fidelity $F_{\rm min}\sim 0.91$ (see Appendix~\ref{appendix: theta state generation}). We can also add the influence of imperfect coupling. This can be done by considering a coupling efficiency $\beta$ to the collection channel. With imperfect coupling the minimal fidelity  $F_{\rm min} \sim 1 - 2(1-0.954\beta)$. 

To increase the fidelity of the deterministically generated $\theta$ state  we propose using the process of $\textit{modal concentration}$. A Takagi decomposition expresses the two-photon wavefunction $\psi_2(x_1,x_2)$ as a sum of a product of separable functions $\phi_i(x)$ and therefore as a sum of two-photon Fock states occupying different modes. This combination can be concentrated into a single Fock state by passing it through a QPG similar to that shown in Fig.~\hyperref[fig:schematic]{1(b)}. The QPG absorbs the dominant two-photon Fock state and the remainder of the state passes to a detector. The absence of a click heralds a successful purification process, while any click on the detector indicates failure and the state is discarded. This process therefore converts the infidelity of the state into heralded errors. In principle, this produces an ideal state. Experimentally, QPGs have been developed with fidelity of 96\% when operating with single photons \cite{Serino2023PRX}.

Besides generating the $\theta$ states, the squeezing operation is another key resource in our protocol. Indeed performing in-line squeezing of a single propagating mode without squeezing the vacuum of other propagating modes is a challenging task and is still an outstanding goal in quantum optics. Nevertheless, the squeezing operation has been experimentally demonstrated on single photons \cite{Miwa2014PRL}. We believe our work shows that an in-line squeezer is a key tool to efficiently produce non-Gaussian states and will therefore encourage further work to develop in-line squeezing protocols.

\section{Discussion and Conclusions}
We propose using squeezed superpositions of zero-photon and two-photon states for generating cat states. When generating cat states of size $|\alpha|^2=4$, the advantage of using the $\theta$ states compare with Gaussian states becomes apparent in the success probabilities, at approximately $0.5$ versus $10^{-2}$ \cite{furusawa,Hanamura2025,molmer_photon_subtraction} while maintaining high fidelities (approximatelly $0.99$). On the other hand, Winnel \textit{et al.}~\cite{tim_ralph_cat} claimed to achieve unit success probabilities with large-photon (up to 20-photon) Fock state inputs, but due to their large linear optical network, small (at 2\% loss) detector inefficiencies compound and significantly reduce the fidelities (to approximately 0.6). In contrast, our proposed linear subtraction protocol can achieve fidelities of 0.98 with similar detection loss. Therefore, we anticipate that cat states generated with our protocols can be used as building blocks for generating larger non-Gaussian states via breeding protocols \cite{Sychev2017Npho,Terhal2018PRA}. Our protocols have the potential to be integrated into state-generation protocols for error-corrected quantum computing with GKP states \cite{furusawa_gkp, Bourassa2021Quantum}.

\section*{Acknowledgments} S.M. acknowledges support from the Australian Research Council (ARC) via the Future Fellowship, `Emergent many-body phenomena in engineered quantum optical systems', project no. FT200100844. H.L. acknowledges support from the Sydney Quantum Academy, Sydney, NSW, Australia.
\bibliography{main}

\newpage
\appendix
\onecolumngrid

\section{Squeezed $\ket{\theta}$ state and the mean photon number}
\label{appendix:theta state}
\begin{figure}[h!]
  \includegraphics[width=0.5\columnwidth]{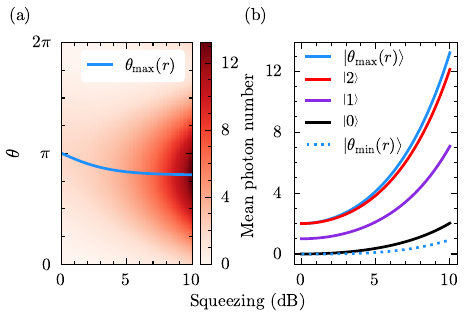}
  \caption{Mean photon number of few-photon states under squeezing. (a) Mean photon number of the $\ket{r,\theta}$ state. The blue line highlights $\theta_{\text{max}}(r)$. (b) Mean photon number of $\theta_{\rm max}(r)$ (solid blue line), $\theta_{\rm max}(r)$ (dotted blue line), vacuum (black line), single-photon (purple line) and two-photon (red line) states.}
  \label{fig:mean_pho_num}
\end{figure}

In this appendix, we calculate the mean photon number and $\theta_{\rm max}(r)$ showcasing the flexibility of the $\theta$ states under squeezing compared with Fock states. We start with the squeezing operator identities,
\begin{align}
    \hat{S}^\dagger(r)\hat{a}\hat{S}(r)=\hat{a}\cosh r+\hat{a}^\dagger\sinh r,\quad
    \hat{S}^\dagger(r)\hat{a}^\dagger\hat{S}(r)=\hat{a}^\dagger\cosh r+\hat{a}\sinh r.
\end{align}
The mean photon number is expressed as
\begin{align}
    \bra{\theta}\hat{S}^\dagger(r)\hat{n}\hat{S}(r)\ket{\theta}=\cosh(2r)\left(\frac{3}{2}-\cos(\theta)\right)+\frac{\sinh(2r)}{\sqrt{2}}\sin(\theta)-\frac{1}{2}
\end{align}
The maximum and minimum are
\begin{align}
    \theta_{\rm max}(r)=\pi-\arctan(\frac{\tanh(2r)}{\sqrt{2}}), \quad
    \theta_{\rm min}(r)=2\pi-\arctan(\frac{\tanh(2r)}{\sqrt{2}})
\end{align}
for $\theta\in[0,2\pi)$. The mean photon number with $\theta_{\rm max(+)}(r)$  and $\theta_{\rm min(-)}$ is expressed as
\begin{align}
    \cosh(2r)\left(\pm\sqrt{\frac{2}{\tanh^2(2r)+2}}+\frac{3}{2}\right)\pm\frac{\sinh(2r)\tanh(2r)}{\sqrt{2(\tanh^2(2r)+2)}}-\frac{1}{2}.
\end{align}
In Fig.~\ref{fig:mean_pho_num}, we compare the mean photon number of the squeezed $\theta$ states with that of the vacuum, single-photon and two-photon states. If we choose $\theta=\theta_{\rm max}(r)$, the squeezed $\theta$ states have larger mean photon numbers than the squeezed two-photon states, while choosing $\theta=\theta_{\rm min}(r)$ gives a smaller average photon number than a squeezed vacuum. This highlights the flexibility of the $\theta$ states under squeezing.

\section{Linear photon subtraction}
\label{appendix: linear photon subtraction}

Here we provide the details of the linear photon subtraction protocol using a beamsplitter. Particularly, we calculate the single-photon-subtracted output state wavefunction and the stellar rank using continuous variables. The many-photon subtracted output states while considering detector inefficiencies are calculated using discrete variables. Additionally, we demonstrate the performance of the $\theta$ states by comparing to that of the single and two-photon states inputs (Fig.~\ref{fig:linear compare}). Lastly, we consider the output state after many-photon subtraction events by showing their Wigner functions (Fig.~\ref{fig:linear wigners}), and surprisingly, the two-photon subtracted state is also a high quality even-cat state (Fig.~\ref{fig:linear 1 and 2 clicks}).

\subsection{Continuous variable calculation}
To calculate the exact analytic continuous variable position wavefunction of the single-photon-subtracted output state, we first conveniently transform the beamsplitter unitary, parametrised by $\gamma$, to real coefficients by transforming the second mode operators $i\hat{a}_2\rightarrow\hat{a}_2$ and $i\hat{b}_2\rightarrow\hat{b}_2$. Effectively, the input and output modes are now related by
\begin{align}
    \begin{pmatrix}
        \hat{a}_1\\\hat{a}_2
    \end{pmatrix}=
    \begin{pmatrix}
        \cos(\gamma) & -\sin(\gamma)\\
        \sin(\gamma) & \cos(\gamma)
    \end{pmatrix}
    \begin{pmatrix}
        \hat{b}_1\\\hat{b}_2
    \end{pmatrix}.
\end{align}
Since the squeezing operator and the operator to create a $\theta$ state from a vacuum consist of even powers of the annihilation and creation operators, effectively the squeezing strength and angle are transformed into $r_2\rightarrow-r_2$ and $\theta_2\rightarrow-\theta_2$. We now define the position and momentum operators of the output modes $\hat{x}_j=1/\sqrt{2}(\hat{b}_j+\hat{b}_j^\dagger)$ and $\hat{p}_j=i/\sqrt{2}(\hat{b}_j^\dagger-\hat{b}_j)$, where $j=1,2$ denotes the output modes. Under the action of the beamsplitter, the output position variables $(x_1,x_2)$ are related to the input position variables $(x_1',x_2')$ by
\begin{align}
    \begin{pmatrix}
        x'_1\\x'_2
    \end{pmatrix}=
    \begin{pmatrix}
        \cos(\gamma) & \sin(\gamma)\\
        -\sin(\gamma) & \cos(\gamma)
    \end{pmatrix}
    \begin{pmatrix}
        x_1\\x_2
    \end{pmatrix}.
\end{align}

The wavefunction for the squeezed vacuum state is expressed as \cite{CV_review}
\begin{equation}
    \psi(x)=\pi^{-1/4}s^{-1/2}e^{-x^2/2s^2},
\end{equation}
where $s=e^r$. Hence, the wavefunction for a squeezed $\theta$ state is expressed as
\begin{equation}
    \Psi_i(x)=\left[\cos\left(\frac{\theta_i}{2}\right)+\sin\left(\frac{\theta_i}{2}\right)\left(\frac{\sqrt{2}x^2}{s_i^2}-\frac{1}{\sqrt{2}}\right)\right]\psi(x),
\end{equation}
where $i=1,2$ denotes the output modes. Projection of one of the output modes, parametrised by $x_2$, onto the single-photon Fock state yields the unnormalised wavefunction
\begin{align}
    \psi_{\rm out}(x_1)=\int_{-\infty}^\infty dx_2~\Psi_1(\cos\gamma x_1+\sin\gamma x_2)\Psi_2(-\sin\gamma x_1+\cos\gamma x_2)\psi_1(x_2),
\end{align}
where $\psi_1(x)=\sqrt{2}\pi^{-1/4}xe^{-x^2/2}$ is the single-photon Fock state wavefunction. Focusing on the exponential terms and completing the square, we obtain
\begin{align}
    &\exp[-(\cos\gamma x_1+\sin\gamma x_2)^2/2s_1^2-(-\sin\gamma x_1+\cos\gamma x_2)^2/2s_2^2-x_2^2/2]\\
    &=\exp[-a(x_2+bx_1/a)^2]\exp[-cx_1^2],
\end{align}
where
\begin{equation}
    a=[\sin^2(\gamma)/s_1^2+\cos^2(\gamma)/s_2^2+1]/2,\quad b=(1/s_1^2-1/s_2^2)\sin(2\gamma)/4,\quad c=[\cos^2(\gamma)/s_1^2+\sin^2(\gamma)/s_2^2-2b^2/a]/2.
\end{equation}
This form allows us to integrate the expression using the relation
\begin{align}
    \int_{-\infty}^\infty dx ~x^ne^{-\alpha(x+\beta)^2}=\sum_{k=0}^{\lfloor n/2\rfloor}{n \choose 2k}(-\beta)^{n-2k}\sqrt{\frac{\pi}{\alpha}}\frac{(2k-1)!!}{2^k\alpha^k},
    \label{gaussian integral}
\end{align}
for $\alpha>0$ and $\beta\in\mathbb{R}$.

After the beamsplitter, the unnormalized position wavefunction, conditioned on detecting a single photon in the output mode with annihilation operator $\hat{b}_2$, is expressed as
\begin{align}
    \psi_{\rm out}(x_1)=\pi^{-1/4}\sqrt{\frac{2}{ s_1 s_2 a}}e^{-cx_1^2}(b_1x_1+b_3x_1^3+b_5x_1^5)
\end{align}
where
\begin{align}
    b_1=&\frac{1}{a}\left[-Ab+\frac{B-C}{2}\sin(2\gamma)-\frac{3b}{2a}(B\sin^2(\gamma)+C\cos^2(\gamma))+\frac{3}{8a}D\sin(4\gamma)-\frac{15b}{16a^2}D\sin^2(2\gamma)\right],\\
    b_3=&-\frac{b}{a}(B\cos^2(\gamma)+C\sin^2(\gamma))+\frac{D\sin(4\gamma)}{2a}\left(\frac{3b^2}{a^2}-\frac{1}{2}\right)+\frac{b^2}{a^2}(B-C)\sin(2\gamma)-\frac{b^3}{a^3}(B\sin^2(\gamma)+C\cos^2(\gamma))\\
    &-\frac{3b}{2a^2}D\left(\cos(4\gamma)+\frac{\sin^2(2\gamma)}{2}\right)-\frac{5b^3}{4a^4}D\sin^2(2\gamma),\\
    b_5=&\frac{Db}{a}\left[-\left(\frac{b^4}{4a^4}+\frac{b^2}{2a^2}+\frac{1}{4}\right)\sin^2(2\gamma)+\frac{b}{2a}\left(\frac{b^2}{a^2}-1\right)\sin(4\gamma)-\frac{b^2}{a^2}\cos(4\gamma)\right],
\end{align}
where
\begin{align}
    A=&\left[\cos(\theta_1/2)-\sin(\theta_1/2)/\sqrt{2}\right]\left[\cos(\theta_2/2)-\sin(\theta_2/2)/\sqrt{2}\right],\quad
    B=\sqrt{2}s_1^{-2}\sin(\theta_1/2)\left[\cos(\theta_2/2)-\sin(\theta_2/2)/\sqrt{2}\right],\\
    C=&\sqrt{2}s_2^{-2}\sin(\theta_2/2)\left[\cos(\theta_1/2)-\sin(\theta_1/2)/\sqrt{2}\right],\quad
    D=2s_1^{-2}s_2^{-2}\sin(\theta_1/2)\sin(\theta_2/2).
\end{align}
The success probability, $P_{\rm suc}=\int dx_1~\abs{\psi_{\rm out}(x_1)}^2$ is expressed as
\begin{align}
    P_{\rm suc}=\frac{\sqrt{2}}{s_1s_2a\sqrt{c}}\left[\frac{b_1^2}{4c}+\frac{3b_1b_3}{8c^2}+\frac{15(b_3^2+2b_1b_5)}{64c^3}+\frac{105b_3b_5}{128c^4}+\frac{945b_5^2}{1024c^5}\right].
\end{align}
The wavefunction for a squeezed odd-cat state, $\ket{{\rm Cat}_{\alpha,-}}\propto\hat{S}(r)[\hat{D}(\alpha)-\hat{D}(-\alpha)]\ket{0}$ is expressed as \cite{furusawa}
\begin{align}
    &\psi_{\rm cat}(x)=\frac{e^{\alpha^2}}{\sqrt{N_{\rm cat}}}\left[e^{-(x-\sqrt{2}\alpha s)^2/2s^2}-e^{-(x+\sqrt{2}\alpha s)^2/2s^2}\right],\\
    &N_{\rm cat}=2\left(1-e^{-2\abs{\alpha}^2}\right)\pi^{1/2}se^{(\alpha+\alpha^*)^2/2}.
\end{align}
Hence, the fidelity, $F=\abs{\braket{{\rm cat}_{\alpha,-}}{\rm out}}^2$ is
\begin{align}
    F=&\frac{16\sqrt{\pi}s^2}{N_{\rm cat}P_{\rm suc}s_1s_2a(2s^2c+1)}\abs{e^{\alpha^2/(2s^2c+1)}}^2\\
    &\abs{\left(b_1+\frac{3s^2b_3}{2s^2c+1}+\frac{15s^4b_5}{(2s^2c+1)^2}\right)\frac{\sqrt{2}\alpha s}{2s^2c+1}+\left(b_3+\frac{10s^2b_5}{2s^2c+1}\right)\left(\frac{\sqrt{2}\alpha s}{2s^2c+1}\right)^3+b_5\left(\frac{\sqrt{2}\alpha s}{2s^2c+1}\right)^5}^2.
\end{align}

\subsubsection{Stellar rank}

The stellar rank is defined by the number of zeros of the stellar function, $F^\star(\alpha):=e^{|\alpha|^2/2}\braket{\alpha^*}{\psi_{\rm out}}$ \cite{Chabaud2020PRL}. The overlap with a coherent state $\bra{\alpha^*}$ can be readily calculated in the position basis. The wavefunction of a coherent state $\ket{\alpha}$ is expressed as\cite{CV_review}
\begin{equation}
    \psi_\alpha(x)=\pi^{-1/4}e^{-(\alpha+\alpha^*)^2/4+\alpha^2}e^{-(x-\sqrt{2}\alpha)^2/2}.
\end{equation} 
The overlap with the output state is expressed as
\begin{align}
    \braket{\alpha^*}{\psi_{\rm out}}\propto e^{-(\alpha+\alpha^*)^2/4+\alpha^2/(2c+1)}\int_{-\infty}^\infty dx~(b_1x+b_3x^3+b_5x^5)e^{-(2c+1)[x-\sqrt{2}\alpha/(2c+1)]^2/2}.
\end{align}
Notice that,  for $\alpha\in\mathbb{R}$, this is exactly the form of the integral given by Eq.~\eqref{gaussian integral} and the solution is an $n$-order polynomial of $\alpha$. Moreover, for $\alpha\in\mathbb{C}$, one can analytically extend the domain of the finite-order polynomial to the complex plane and find the stellar function
\begin{equation}
        F^\star(\alpha)\propto e^{\abs{\alpha}^2/2-\frac{(\alpha+\alpha^*)^2}{4}+\frac{\alpha^2}{2c+1}}\left[\left(b_1+\frac{3b_3}{2c+1}+\frac{15b_5}{(2c+1)^2}\right)\frac{\sqrt{2}\alpha}{2c+1}+\left(b_3+\frac{10b_5}{2c+1}\right)\left(\frac{\sqrt{2}\alpha}{2c+1}\right)^3+b_5\left(\frac{\sqrt{2}\alpha}{2c+1}\right)^5\right].
\end{equation}
This expression is very general and reduces in several limiting cases. We first see that, in general, the stellar rank is 5. When the squeezing strengths are equal ($b=0$), or one of the input state is in vacuum ($D=0$), $b_5=0$ and the stellar rank is 3. When both conditions are simultaneously satisfied ($b=0$, $D=0$), the coefficients are $b_5=b_3=0$ and the stellar rank is 1. Moreover, if both input states are vacuum and subject to equal squeezing strengths, $b_1=0$, and the stellar rank is 0, in other words, the output state is Gaussian. However, the probability of generating such states is also 0. Nonetheless, if both inputs are vacuum but under different squeezing strengths ($b_3=b_5=0$), the stellar rank is 1. When the input states are fixed to two-photon Fock states, in general, the stellar rank is still 5 but offers less flexibility for tuning the output wavefunction compared with the $\theta$ states. 

\subsection{Discrete variable calculation}

We now present the details of our calculations for linear photon subtraction using a discrete-variable approach. This is particularly useful when we are considering imperfect photon detection via photon loss. To obtain an analytic expression of the many-photon-subtracted output state, first we express the squeezed $\theta$ states in the number basis,
\begin{align}
    \ket{r,\theta}=\hat{S}(r)\ket{\theta}=\sum_{n=0}^\infty B_n(r,\theta)\ket{2n},
\end{align}
where
\begin{align}
\label{Bn}
    B_n(r,\theta)=\frac{\tanh^n(r)}{\sqrt{\cosh(r)}}\frac{\sqrt{(2n)!}}{2^nn!}\left(A_0+\frac{2n}{\tanh(r)}A_++(2n+1)\tanh(r)A_-+2nA_{+-}\right),
\end{align}
where
\begin{align}
    &A_0=\cos\left(\frac{\theta}{2}\right)-\sin\left(\frac{\theta}{2}\right)\frac{\sinh(2r)}{2\sqrt{2}},\quad A_+=\sin\left(\frac{\theta}{2}\right)\frac{\cosh(2r)+1}{2\sqrt{2}}\\
    &A_-=\sin\left(\frac{\theta}{2}\right)\frac{\cosh(2r)-1}{2\sqrt{2}},\quad A_{+-}=-\sin\left(\frac{\theta}{2}\right)\frac{\sinh(2r)}{\sqrt{2}}.
\end{align}
Now consider a beamsplitter parametrised by $\gamma$ and with the input modes, denoted by their annihilation operators $(\hat{a}_1,\hat{a}_2)$. Under conjugation of the two-mode beamsplitter unitary $\hat{U}_{1,2}(\gamma)$, the input modes are linear transformations of the output modes $(\hat{b}_1,\hat{b}_2)$ via
\begin{align}
    \hat{U}_{1,2}(\gamma)\begin{pmatrix}
    \hat{a}^\dagger_1\\
    \hat{a}^\dagger_2
    \end{pmatrix}\hat{U}_{1,2}^\dagger(\gamma)=\begin{pmatrix}
        \cos(\gamma) & i\sin(\gamma)\\
        i\sin(\gamma) & \cos(\gamma)
    \end{pmatrix}
    \begin{pmatrix}
    \hat{b}^\dagger_1\\
    \hat{b}^\dagger_2
    \end{pmatrix}.
\end{align}
Hence, the input states transforms as
\begin{align}
    \hat{U}_{1,2}(\gamma)\ket{r_1,\theta_1}\ket{r_2,\theta_2}&=\sum_{n,m=0}^{\infty}B_n(r_1,\theta_1)B_m(r_2,\theta_2)\frac{(c_\gamma\hat{b}^\dagger_1+s_\gamma\hat{b}^\dagger_2)^{2n}}{\sqrt{(2n)!}}\frac{(s_\gamma\hat{b}^\dagger_1+c_\gamma\hat{b}^\dagger_2)^{2m}}{\sqrt{(2m)!}}\ket{0}\\
    &=\sum_{n,m=0}^\infty B_n(r_1,\theta_1)B_m(r_2,\theta_2)\sum_{k=0}^{2n}\sum_{l=0}^{2m}{2n \choose k}{2m \choose l}\frac{c_\gamma^{2n-k+l}s_\gamma^{k+2m-l}\hat{b}_1^{\dagger 2n-k+2m-l}\hat{b}_2^{\dagger k+l}}{\sqrt{(2n)!(2m)!}}\ket{0},
\end{align}
where $c_\gamma=\cos(\gamma)$ and $s_\gamma=i\sin(\gamma)$. Subtraction of $q$ photons in the context of a beamsplitter is realised by projection of the output mode $\hat{b}_2$ onto the number state $\ket{q}$, and consequently, the unnormalised output state in the mode $\hat{b}_1$ is expressed as
\begin{align}
    &\bra{q}\otimes\mathbf{1}\hat{U}_{1,2}(\gamma)\ket{r_1,\theta_1}\ket{r_2,\theta_2}\\
    =&\sum_{n=0}^\infty\sum_{m=\text{max}(0,\lceil q/2\rceil-n)}^\infty B_n(r_1,\theta_1)B_m(r_2,\theta_2)\sum_{k=\text{max}(0,n-\lfloor q/2\rfloor)}^{2n+\text{min}(0,n-\lceil q/2\rceil)}{2n \choose k}{2m \choose 2n-k+2m-q}\\
    &c_\gamma^{2k+2n-q}s_\gamma^{2(2n-k)+2m-q}\sqrt{\frac{(2n+2m-q)!q!}{(2n)!(2m)!}}\ket{2n+2m-q}.
\end{align}

\begin{figure}[!t]
    \includegraphics[width=0.7\textwidth]{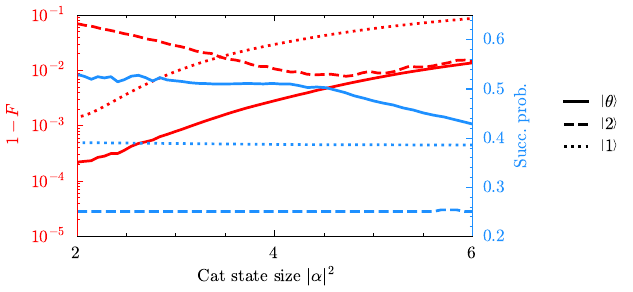}
    \caption{Infidelity (red lines) and success probability (blue lines) for linear photon subtraction. For prioritisation of the success probability together with the fidelity, the squeezed $\theta$ states (solid lines) outperform the squeezed two-photon (dashed lines) and single-photon (dotted) states as inputs, for both criteria. When using the squeezed single-photons, the protocol is heralded by detection of two photons (see the text).}
    \label{fig:linear compare}
\end{figure}

We model inefficient detectors by placing a beamsplitter with transmission coefficient $\cos(\phi)$ before the ideal detector; the detector efficiency is $\epsilon=\cos^2(\phi)$. Again, projecting on the mode $\hat{b}_2$ onto the number state $\ket{q}$ and tracing over the reflected mode, denoted $\hat{b}_3$, gives us the unnormalised output density matrix
\begin{align}
    \hat{\rho}=\sum_{p=0}^\infty\bra{\rm out}\ket{q}\ket{p} \bra{p}\bra{q}\ket{\rm out},
\end{align}
where $\ket{p}=\hat{b}^{\dagger p}_3/\sqrt{p!}\ket{0}$ and $\ket{q}=\hat{b}^{\dagger q}_2/\sqrt{q!}\ket{0}$. The output pure state is
\begin{align}
    \ket{\rm out}=\hat{U}_{2,3}(\phi)\hat{U}_{1,2}(\gamma)\ket{r_1,\theta_1}\ket{r_2,\theta_2}\ket{0},
\end{align}
where $\hat{U}_{2,3}(\phi)$ is the two-mode beamsplitter unitary acting on the modes with annihilation operators $\hat{b}_2$ and $\hat{b}_3$.
Each term in the sum can be calculated by
\begin{align}
    \bra{p}\bra{q}\ket{\rm out}=c_\phi^qs_\phi^p\sum_{n=0}^\infty\sum_{m=\text{max}(0,\lceil (p+q)/2\rceil-n)}^\infty B_n(r_1,\theta_1)B_m(r_2,\theta_2)\sqrt{\frac{(2n+2m-p-q)!p!q!}{(2m)!(2n)!}}\ket{2n+2m-p-q}\\
    \sum_{k=\max(0,p+q-2m)}^{\min(p+q,n+\lfloor(p+q)/2\rfloor,2n)}{2n \choose k}{2m \choose p+q-k}c_\gamma^{2n-2k+p+q}s_\gamma^{2m+2k-p-q}\sum_{l=\max(0,k-q)}^{p}{k \choose l}{p+q-k \choose p-l},
\end{align}
where $c_\phi=\cos(\phi)$ and $s_\phi=i\sin(\phi)$.

From Fig.~\ref{fig:linear compare}, the $\theta$ states outperform the single-photon and two-photon states with lower infidelities and higher success probabilities. The improvement of the success probability by a factor of 2 over two-photon inputs is significant. This is because, for many applications, high-photon-number cat states are required \cite{tim_ralph_quantum_computation,Bergmann2016PRA,Li2017PRL,Xu2023npj,Schlegel2022PRA,Le2018PRL,Grimsmo2020PRX,Chamberland2022PRXQ,Gravina2023PRXQ,Lee2024PRXQ,tim_ralph_cat,Albert2016PRL}. These larger cat states are prepared with probabilistic linear optics circuits \cite{Sychev2017Npho,amplify_large_cat,amplify_large_cat_2}, also known as ``enlargement" or cat state ``breeding", where two cat states are combined to generate a cat state with size $|\alpha_{\rm new}|^2=2|\alpha|^2$. For example, the generate of a cat state with $|\alpha|^2\approx 20$, without the use of additional resources of delays and switching, requires two iterations of the cat state ``breeding" protocols and four input cat states with size $|\alpha|^2\approx5$. The total probability therefore scales as $P_{\rm suc}^4$, and the factor-of-2 improvement in the input cat state preparation is then amplified to a factor-of-16 improvement in the total success probability. We also note that, when single-photon states are used as inputs, the protocol heralds on two-photon subtractions. This is because a trivial beamsplitter will return unit success probability, and the generated cat state is then poorly approximated by a squeezed single-photon state.

\subsection{Multiphoton-subtracted states}
\begin{figure}[h]
    \centering
    \includegraphics[width=0.75\linewidth]{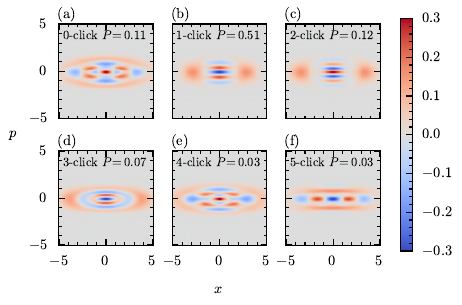}
    \caption{The squeezed $\theta$ states are used as inputs for the beamsplitter. (a)-(f) Wigner functions of the output state, conditioned on detecting zero to five photons in one of the output modes. $P$ is the success probability of the corresponding detection event. The parameters in all panels, $r_1$, $r_2$, $\theta_1$, $\theta_2$, $\gamma$ and the post unsqueezing strength $r_{\rm post}$ are optimised for (b), the single-photon-subtraction output state for producing cat state of size $\abs{\alpha}^2=4$. Remarkably, under the same parameters, the two-photon-subtracted state in (c), is a high-quality even-cat state of size $\abs{\alpha}^2\approx5.9$.}
    \label{fig:linear wigners}
\end{figure}

We now consider the output state after multiphoton subtraction events via the beamsplitter and using squeezed $\theta$ states as inputs. In Fig.~\ref{fig:linear wigners} we plot the Wigner functions for zero-photon-subtracted to five-photon-subtracted output states, under the same parameters optimised for the single-photon subtracted protocol. Remarkably, the two-photon subtracted state is a high quality even-cat state.

\begin{figure}[h]
    \centering
    \includegraphics[width=0.7\linewidth]{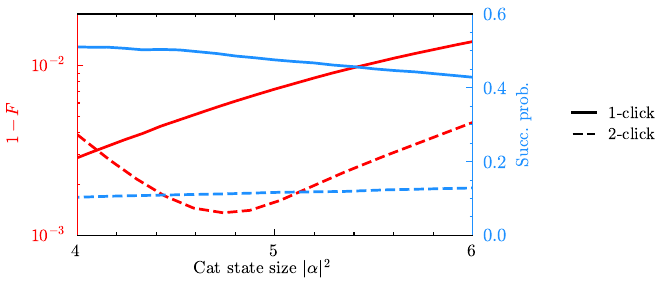}
    \caption{Squeezed $\theta$ states are used as inputs for the beamsplitter. The infidelity (red) and the success probability (blue) for single-photon (solid) and two-photon (dashed) subtracted output states. The two-photon subtracted state uses the same input parameters as the single-photon subtracted state but the cat state size $\abs{\alpha}^2$ and the post unsqueezing strength are further optimised for the smallest infidelities. The two-photon subtracted state can only produce cat state size of $\abs{\alpha}^2\geq4$.}
    \label{fig:linear 1 and 2 clicks}
\end{figure}

Furthermore, in Fig.~\ref{fig:linear 1 and 2 clicks}, we compare the infidelity and the success probability of the single-photon-subtracted and two-photon-subtracted output. Under the same input parameters, the two-photon-subtracted output state can only produce cat states of size $\abs{\alpha}^2\geq4$ but has smaller infidelities than that of the single-photon subtracted case. However, the success probabilities are much lower at approximately $0.12$.

\section{Nonlinear photon subtraction}
\label{appendix: nonlinear photon subtraction}
In this appendix, we provide more details on the nonlinear photon subtraction protocol and the photon field dynamics in the MPS formalism. First, we introduce the cascaded system consisting of a virtual input cavity, the TLS and a virtual capture cavity. The virtual cavities are used to insert a single mode into and collect a single mode from the one-dimensional photon field. We then write the time-evolution unitary as a solution to the quantum stochastic Schr\"{o}dinger equation. By discretizing the waveguide into small time bins, we show that the system and the photon field pure state can be written as an MPS. Time evolution is then generated by applying local operators to the MPS. Finally, we compare the performance of the squeezed $\theta$ states with that of the squeezed single-photon states used as inputs.

We consider a one-dimensional waveguide with a TLS chirally coupled to a right-propagating photon field. We include two virtual cavities to model the release and capture of photons in specific temporal modes. Under the Born-Markov approximation and the rotating-wave approximation together with renormalisation to the resonance frequency of the TLS, the Hamiltonian for this system is ($\hbar=v_g=1$)
\begin{align}
\label{full_hamiltonian}
    \hat{H}=&-i\int_{\mathbb{R}} dx \hat{a}^\dagger(x)\partial_x\hat{a}(x)+\sqrt{\Gamma}[\hat{a}^\dagger(x_1)\hat{\sigma}_-+\hat{\sigma}_+\hat{a}(x_1)]+\hat{H}_{\text{in}}(t)+\hat{H}_{\text{cap}}(t),
\end{align}
where $\hat{a}^\dagger(x)$ [$\hat{a}(x)$] is the photon creation (annihilation) operator at position $x$, satisfying the commutation relation $\comm{\hat{a}(x)}{\hat{a}^\dagger(x')}=\delta(x-x')$, and $\hat{\sigma}_+$ and $\hat{\sigma}_-$ are the Pauli operators of the TLS. The first term in the Hamiltonian describes the right-propagating photon field while the second term describes the interaction between the TLS and the photon field at position $x_1$ with coupling strength $\Gamma$  \cite{quantum_chiral_optics,bound_states_dynamics}. To include a general input photon state and to analyse a specific mode in the photon field, we include two Hamiltonians with dynamical coupling strengths, responsible for the release and capture of modes with complex wave packets $u(t)$ and $v(t)$ into and out of the waveguide \cite{input_output_quantum_pulses}: $\hat{H}_{\text{in}}(t)=g_u^*(t)\hat{a}^\dagger(x_0)\hat{c}_{\text{in}}+g_u(t)\hat{c}^\dagger_{\text{in}}\hat{a}(x_0)$ and $\hat{H}_{\text{cap}}(t)=g_v^*(t)\hat{a}^\dagger(x_2)\hat{c}_{\text{cap}}+g_v(t)\hat{c}^\dagger_{\text{cap}}\hat{a}(x_2)$, where $\hat{c}^\dagger_{\text{in}}$ $(\hat{c}_{\text{in}})$ and $\hat{c}^\dagger_{\text{cap}}$ $(\hat{c}_{\text{cap}})$ are the excitation creation (annihilation) operators for the input and capture virtual cavities, respectively. 

The full Hamiltonian in Eq. \eqref{full_hamiltonian} describes a cascaded system where a vacuum field sequentially interacts with the input cavity, the TLS, and the capture cavity, i.e., the corresponding positions satisfy $x_0<x_1<x_2$. Here, we proceed to an interaction frame with respect to the right-propagating photon field in the absence of the chiral atom. This is equivalent to transforming the argument of the photon field operators to $\hat{a}(x)\rightarrow\hat{a}(x-t)$. Now we can label the photon field operators using $t$, and they satisfy the commutation relation $\comm{\hat{a}(t)}{\hat{a}^\dagger(t')}=\delta(t-t')$. We emphasise that this commutation relation is the result of taking $v_g=1$ and moing into the interaction frame, and should not be confused with that of the Heisenberg picture. Consequently, a right-propagating mode creation operator for a normalised complex wave packet $u(t)$ reads
\begin{equation}
    \hat{a}^\dagger_u=\int_0^\infty dt\,u(t)\hat{a}^\dagger(-t).
\end{equation}
The time-dependent coupling strength is chosen as
\begin{equation}
    g_{u}(t)=\frac{-iu^*(t)}{\sqrt{1-\int_0^tdt'\,\abs{u(t')}^2}},
\end{equation}
such that the input cavity initialised at $t=0$ will asymptotically release a specific state into the waveguide in a temporal mode $u(t)$ \cite{input_output_quantum_pulses,input_mode,flying_quantum_bits}. Similarly, the capture cavity will asymptotically capture photons in a temporal mode $v(t)$ if the coupling strength is chosen to be \cite{capture_mode}
\begin{equation}
\label{capture}
    g_v(t)=\frac{iv^*(t)}{\sqrt{\int_0^tdt'\,\abs{v(t')}^2}}.
\end{equation}

The photon field dynamics and the nonlinear photon subtraction protocol are described as follows. At $t=0$, the input cavity is prepared in the state $\ket{r,\theta_{\rm max}}$. Release of the resource state into a Gaussian mode  $u(t)=\frac{1}{\sqrt{\sigma}\pi^{1/4}}e^{-(t-t_0)^2/2\sigma^2}$ of width $\sigma$ is generated by the interactions between the input cavity and the vacuum field. The many-photon cavity interaction is modeled numerically by solving a quantum stochastic Schr\"{o}dinger equation using the MPS description, explained in the next section. In terms of the mode creation operators, the state of the photon field is expressed as
\begin{equation}
\label{input state}
    \sum_{n=0} B_n(r,\theta_{\rm max})\frac{\hat{a}^{\dagger 2n}_u}{\sqrt{(2n)!}}\ket{0},
\end{equation}
where $B_n(r,\theta)$ are defined by Eq.~\eqref{Bn}. After scattering with the TLS (i.e., for a sufficiently long time, the input cavity and the TLS both reach their ground states), we calculate the first-order coherence $G^{(1)}(t,t')=\expval{\hat{a}^\dagger(t)\hat{a}(t')}=\sum_i\Bar{n}_iv^*_i(t)v_i(t')$ and diagonalise it to find the dominant mode $v_1(t)$, where $\Bar{n}_1\geq\Bar{n}_i$ for all $i$. We extract the dominant mode $v_1(t)$ using a virtual cavity by choosing $g_{v_1}(t)$ according to Eq.~\eqref{capture}. To achieve maximum fidelity, i.e., maximum overlap with a pure target state, one must also maximise the purity of the prepared state. But now the nondominant modes are entangled with the virtual cavity, and the purification is performed via a single-photon subtraction by tracing over all single-photon detection events across all possible detection times (i.e., projection onto the single-photon sector). The output state collected by the virtual cavity is generally entangled with the multimode output photon field. For example, for two modes (a dominant mode in the cavity and an ancillary mode in the field), upon projective measurement on the ancillary mode, the purity of the state in the dominant mode would be unity. 

\begin{figure}[t]
    \centering
    \includegraphics[width=0.5\linewidth]{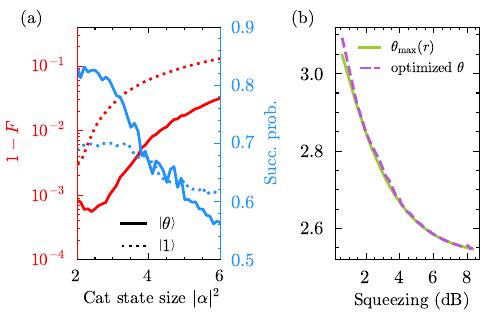}
    \caption{(a) Infidelity (red lines) and success probability (blue lines) for the nonlinear photon subtraction protocol. Squeezed $\theta$ states (solid lines) and the squeezed single-photon states (dotted lines) are used as inputs. (b) $\theta_{\rm max}(r)$ (solid green line) and optimised $\theta$ (dashed purple line) vs the initial squeezing strength.}
    \label{fig:non-linear sup}
\end{figure}

We adopt a gradient-descent optimisation algorithm \cite{adam} to optimise the squeezing strength and the wave packet width for maximum fidelity against a given cat state of size $\abs{\alpha}^2$. The optimised output state is occupied by approximately two modes, $(\Bar{n}_1+\Bar{n}_2)/\sum_i\Bar{n}_i\sim0.95$, depending on the squeezing strength. Furthermore, in Fig.~\hyperref[fig:non-linear sup]{8(b)} we show that $\theta_{\rm max}(r)$ matches the optimised $\theta$ if it was included as an optimisation parameter. The choice of $\theta_{\rm max}$ can be qualitatively explained with a $\pi$-pulse model, $2\int dt \sqrt{n\Gamma}u(t)=\pi$. For high-photon-Fock states, the $\pi$-pulse model well approximates the optimal pulse width for single-photon subtractions \cite{molmer_photon_subtraction}, $\sigma_{\rm opt}=\pi^{3/2}/8n\Gamma$. On the other hand, subtraction from a superposition state, $\sum_n c_n\ket{n}$, requires careful choices for the optimal pulse width. Since the vacuum cannot participate in photon subtraction, for small squeezing strengths $r$, the optimal strategy is to minimise the amplitude of the vacuum. The angle that guarantees zero overlap with the vacuum is $\theta_0=2\arctan\left(\sqrt{2}/\tanh r\right)$, and $\theta_{\rm max}$ agrees with that up to the second order in the Taylor expansion around $r=0$:
\begin{align}
    \theta_{\rm max}\approx\theta_0=\pi-\sqrt{2}r+\order{r^3}
\end{align}
However, for large squeezing strengths, the pulse width is optimal if the distribution of the amplitudes $c_n$ covers the region with slowly varying optimal widths $\sigma_{\rm opt}$, i.e., minimising $\abs{\partial_n\sigma_{\rm opt}}\propto 1/n^2$. Hence, for a fixed squeezing strength, $\theta_{\rm max}$ precisely minimises $1/\expval{n}^2$. The inefficiencies of occupying the vacuum and the few-photon Fock states are compensated by occupation of larger Fock states with slowly varying Fock state optimal widths.

\begin{figure}[!t]
    \centering
    \includegraphics[width=0.5\linewidth]{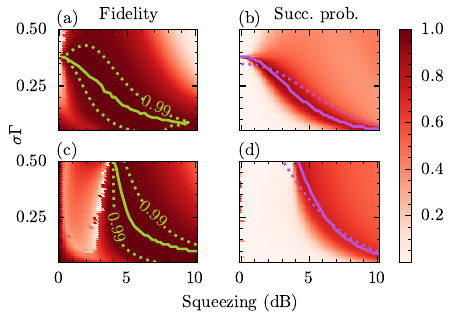}
    \caption{(a),(c) Fidelity and (b),(d) success probability with use of the (a),(b) $\theta$ states and (c),(d) single-photon states as inputs for the nonlinear photon subtraction protocol. In all panels, the input states are subject to initial squeezing strength up to 10~dB and are released into the waveguide in a Gaussian wave packet of width $\sigma$. Moreover, the highest fidelity (success probability) across all squeezing strengths is highlighted by the solid green (purple) line. In (a),(c), the region with fidelity $F>0.99$ is enclosed by the dotted green contour. In (b),(d), the $\pi$-pulse estimation is given by the dotted purple line.}
    \label{fig:non-linear 4}
\end{figure}

In Fig.~\hyperref[fig:non-linear sup]{8(a)} we compare the performance of the squeezed $\theta$ states and the squeezed single-photon states as inputs. The $\theta$ states have an order-of-magnitude smaller infidelities and much higher success probabilities for generating a cat state of size $\abs{\alpha}^2\leq4$. Additionally, the single photons require approximately 10~dB of squeezing to generate a cat state of size $\abs{\alpha}^2\approx 2$. In Fig.~\ref{fig:non-linear 4}, we showcase the fidelity and success probability in parameter space (squeezing strength and wave packet width) for the squeezed $\theta$ states and single-photon states. Particularly, in Fig.~\hyperref[fig:non-linear 4]{9(d)}, the success probability of detecting a photon in the nondominant mode exhibits a jump at approximately 4~dB of initial squeezing. From Fig.~\hyperref[fig:mean_pho_num]{4(b)}, the mean photon number of the 4~dB squeezed single-photon state is approximately 2. The interpretation is that the nondominant modes need to have sufficient amplitude over the single-photon sector for single-photon subtraction to have high success probability.

\subsection{Matrix product state description}
Here we adopt the MPS description for solving the quantum stochastic Schr\"{o}dinger equation \cite{mps,zoller_mps,time_bin_mps1,time_bin_mps2,time_bin_mps3,time_bin_mps4, Gardiner2004,Combes2017APX} and calculating the first-order coherence and performing single-photon detection on the photon field. We begin by writing the solution to the quantum stochastic Schr\"{o}dinger equation, followed by approximating the total time evolution by a product of first-order-Trotter-decomposed unitary operators on a discretized photon field. Finally, we represent the dynamics of the entire system and the photon field in the MPS formalism.
We start by defining the quantum Wiener process
\begin{equation}
    B(t)=\int_0^tdt'\hat{a}(-t').
\end{equation}
With $dB(t)=B(t+dt)-B(t)$, the Stratonovich quantum stochastic Schr\"{o}dinger equation is
\begin{equation}
    d\hat{U}(t)=-i[g_u^*(t)dB^\dagger(t-x_0)\hat{c}_{\rm in}+\sqrt{\Gamma}dB^\dagger(t-x_1)\hat{\sigma}_-+g_v^*(t)dB^\dagger(t-x_2) \hat{c}_{\rm cap}+{\rm H.c.}]\hat{U}(t),
\end{equation}
with the solution
\begin{equation}
    \hat{U}(t,0)=\mathcal{T}\exp{-i\int_0^t[g_u^*(t')dB^\dagger(t'-x_0)\hat{c}_{\rm in}+\sqrt{\Gamma}dB^\dagger(t'-x_1)\hat{\sigma}_-+g_v^*(t')dB^\dagger(t'-x_2) \hat{c}_{\rm cap}+{\rm H.c.}]},
\end{equation}
where $\mathcal{T}$ denotes the time-ordered exponential. Discretizing the photon field into time bins of width $\Delta t$ and defining the quantum noise increments as
\begin{align}
    \Delta B(t_k)=B((k+1)\Delta t)-B(k\Delta t)=\int_{k\Delta t}^{(k+1)\Delta t}dt~\hat{a}(-t),
\end{align}
and they satisfy the commutation relation $\comm{\Delta B(t_k)}{\Delta B^\dagger(t_{k'})}=\Delta t \delta_{k,k'}$. These quantum noise increments are related to the ladder operators of each time bin by $\hat{a}_k=\Delta B(t_k)/\sqrt{\Delta t}$, where $\comm{\hat{a}_k}{\hat{a}^\dagger_{k'}}=\delta_{k,k'}$. The total time evolution operator can be expressed in steps of $\Delta t$:
\begin{align}
    \hat{U}(t,0)=\lim_{N\rightarrow\infty}\hat{U}(t,N\Delta t)\cdots\hat{U}(\Delta t,0).
\end{align}
This allows us to approximate each evolution time step up to the first order in $\Delta t$
\begin{align}
    \hat{U}((k+1)\Delta t,k\Delta t)&\approx\exp{-i[g_u^*(t_k)\Delta B^\dagger(t_{k})\hat{c}_{\rm in}+\sqrt{\Gamma}\Delta B^\dagger(t_{k-1})\hat{\sigma}_-+g_v^*(t_k)\Delta B^\dagger(t_{k-2}) \hat{c}_{\rm cap}+{\rm H.c.}]},\\
    &=\hat{U}_{k,\rm cap}\hat{U}_{k,\rm TLS}\hat{U}_{k,\rm in}.
    \label{time evolution unitary}
\end{align}
Since the input photon bath and the capture cavity are in vacuum and the TLS is in the ground state, we have conveniently defined $x_0=0$, $x_1=\Delta t$ and $x_2=2\Delta t$. 

The physical interpretation of this unitary is that, the input field (time bin) of the TLS is the output field (time bin) of the input cavity and similarly for the capture cavity and the TLS. In other words, this unitary describes a cascaded system. We can express the unitary as a product of three two-body local unitary operators because each part of the system is coupled to different time bins. Finally, the total time evolution of the entire system and the photon field is generated by applying these unitary operators in a time-ordered fashion.

We now introduce the MPS description of the system and the time bin representation of the photon field. Initially, the time bins are all in a vacuum, the input cavity is in the resource state $\ket{r,\theta_{\rm max}}$, the TLS is in the ground state and the capture cavity is in a vacuum, i.e.,
\begin{equation}
    \ket{\psi(0)}=\ket{0}^{\otimes N}\ket{r,\theta_{\rm max}}_{\rm in}\ket{g}_{\rm TLS}\ket{0}_{\rm cap}.
\end{equation}
However, the general state of the system and time bins can be expressed as
\begin{align}
    \ket{\psi}=\sum_{\{i_n,\cdots ,i_0,s_{\rm in},s_{\rm TLS},s_{\rm cap}\}}\psi_{i_n,\cdots ,i_0,s_{\rm in},s_{\rm TLS},s_{\rm cap}}\ket{i_n,\cdots ,i_0,s_{\rm in},s_{\rm TLS},s_{\rm cap}},
\end{align}
where $\ket{i_k}=\Delta B^{\dagger i_k}(t_k)/\sqrt{\Delta t^{i_k} i_k!}\ket{0}_{t_k}$ is the number state of the $k$th time bin, and $\ket{s_{\rm in},s_{\rm TLS},s_{\rm cap}}$ are the states of the system (the cavities and the TLS), $\psi_{i_n,\cdots ,i_0,s_{\rm in},s_{\rm TLS},s_{\rm cap}}$ is the wavefunction and $\{i_n,\cdots ,i_1,s_{\rm in},s_{\rm TLS},s_{\rm cap}\}$ denotes summing over all configurations. The underlying Hilbert space grows exponentially and is infeasible to track a general state. Here, we adopt the MPS approximation to the wavefunction
\begin{align}
    \ket{\psi}\approx\sum_{\{i_n,\cdots ,i_0,s_{\rm in},s_{\rm TLS},s_{\rm cap}\}}A[i_n]\cdots A[i_0]A[s_{\rm in}]A[s_{\rm TLS}]A[s_{\rm cap}]\ket{i_n,\cdots ,i_0,s_{\rm in},s_{\rm TLS},s_{\rm cap}},
    \label{mps}
\end{align}
where each $A[i_k]$ of the physical index $i_k$, is a matrix of size $D_{i_{k+1}}\times D_{i_k}$ representing the wavefunction of the $k$th time bin. Similarly, the system matrices are adjacent to each other forming the matrix product $A[s_{\rm in}]A[s_{\rm TLS}]A[s_{\rm cap}]$. Hence, including the physical index, each $A[i_k]$ is a rank-3 tensor. At first sight, the advantage of adopting the MPS ansatz is not apparent. However, we can restrict the number of excitations in each time bin to only have a few photons (at most two), which is appropriate in the limit of $\Gamma\Delta t\ll1$, subsequently, reducing the physical dimension of each tensor. Observe that these time bin matrices are intrinsically time ordered, labeled by the time when they interact with the input cavity, $k\Delta t$. Further, the system matrices are also ordered by the time at which they interact with the zeroth time bin. We note that the MPS structure with embedded time ordering product of the matrices gives rise to a natural extension to the continuous matrix product state by taking $\Delta t\rightarrow0$ \cite{cmps,cmps2,cmps3}. 

\begin{figure}[h]
    \centering
    \includegraphics[width=\linewidth]{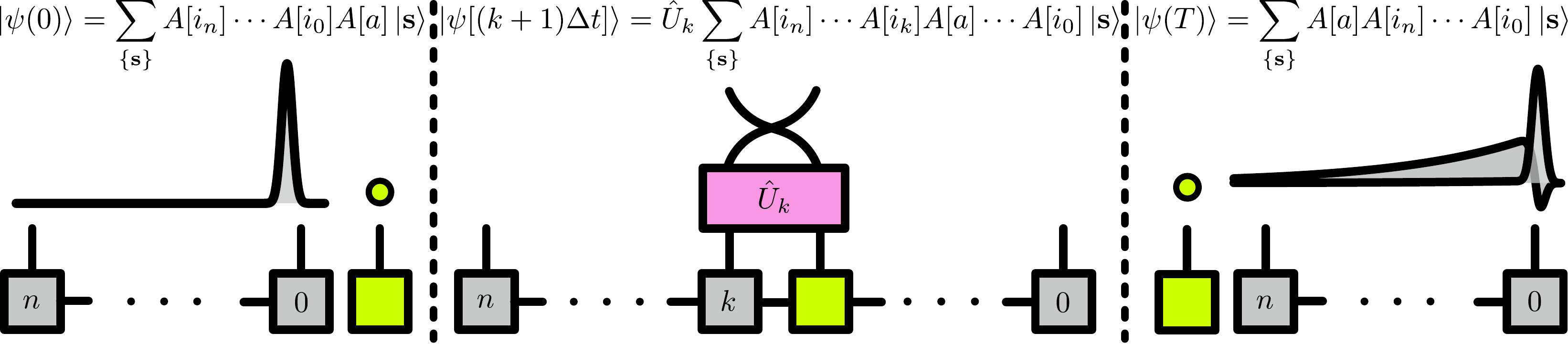}
    \caption{Time evolution of the system. The equations above each panel describe the system and the field at different times, with use of the MPS ansatz. $\{\bold{s}\}$ denotes summing over all configurations of the system and the field. In the left panel, at the initial time, the field is in the discretized state described by Eq.~\eqref{input state} contained in a Gaussian wave packet and the TLS is in its ground state. The time bins are represented by the grey blocks, labeled by the order in which they interact with the TLS (represented by the green block). Moreover, these blocks correspond to the matrices in the MPS (equation above), and the matrix and physical indices are represented by the  horizontal and vertical ``legs" attached to each block, respectively. Hence, matrix multiplication is represented by joining the horizontal ``legs" together. The middle panel depicts time evolution by updating the TLS matrix and the adjacent time bin matrix with use of the unitary operator defined by Eq.~\eqref{mpo}. Performing SVD after swapping  ofthe corresponding physical indices of the time bin and TLS factorises the entire state back to the MPS form and it is now ready for evolution to the next time step. For sufficiently long time, in the last panel, the TLS returns to its ground state and the field is now in a multimode state.}
    \label{fig:mps time evo}
\end{figure}

Time evolution to the next time step is generated by applying a slightly modified unitary operator of Eq.~(\ref{time evolution unitary})
\begin{align}
    \hat{U}'_0\ket{\psi(t_0)}=\ket{\psi(t_1)}=\sum_{\{i_n,\cdots ,i_0,s_{\rm in},s_{\rm TLS},s_{\rm cap}\}}A[i_n]\cdots A[i_1]A[s_{\rm in}]A[s_{\rm TLS}]A[s_{\rm cap}]A[i_0]\ket{i_n,\cdots ,i_0,s_{\rm in},s_{\rm TLS},s_{\rm cap}},
\end{align}
where $\hat{U}'_0=\hat{V}_{\rm swap}\hat{U}_{0,\rm cap}\hat{V}_{\rm swap}\hat{U}_{0,\rm TLS}\hat{V}_{\rm swap}\hat{U}_{0,\rm in}$. The swap operators swap the ordering of the matrices of the interacting parties such that the following unitary is a local two-body unitary, i.e., $A'''[i_0]A'[s_{\rm in}]A'[s_{\rm TLS}]A'[s_{\rm cap}]\rightarrow A[s_{\rm in}]A''[i_0]A'[s_{\rm TLS}]A'[s_{\rm cap}] \rightarrow A[s_{\rm in}]A[s_{\rm TLS}]A'[i_0]A'[s_{\rm cap}]\rightarrow A[s_{\rm in}]A[s_{\rm TLS}]A[s_{\rm cap}]A[i_0]$. After application of each unitary and swap operator, the two-site tensor is factorised by singular value decomposition (SVD) into matrices satisfying the form given by Eq.~\eqref{mps}. Notice that we have omitted the time bins $A[i_{-1}]$ and $A[i_{-2}]$. Since these time bins and the capture cavity are in a vacuum and the TLS is in the ground state, the interaction between these is trivial. Furthermore, the time bin widths should be chosen such that $\Gamma\Delta t\ll1$ and consequently the total time evolution should be complete, i.e., there should be a sufficient number of time bins (tensors) such that the input cavity and the TLS are in their ground states and the capture cavity decouples from the time bins. A diagrammatic example of the time evolution is given in Fig.~\ref{fig:mps time evo}.

We have calculated the matrix-product operator representation of the TLS unitary analytically,
\begin{align}
\label{mpo}
    \hat{U}_{k,\rm TLS}=\begin{pmatrix}-i\sqrt{\Gamma}\Delta B(t_k)\sinc[\sqrt{\Gamma\Delta t\Delta\Lambda(t_k)}]\\
    -i\sqrt{\Gamma}\sinc[\sqrt{\Gamma\Delta t\Delta\Lambda(t_k)}]\Delta B^\dagger(t_k)\\
    \cos[\sqrt{\Gamma\Delta t(\Delta \Lambda(t_k)+1)}]\\
    \cos[\sqrt{\Gamma\Delta t\Delta\Lambda(t_k)}]
    \end{pmatrix}^T\begin{pmatrix}
        \hat{\sigma}_+\\
        \hat{\sigma}_-\\
        \ket{e}\bra{e}\\
        \ket{g}\bra{g}
    \end{pmatrix},
\end{align}
where $\Delta \Lambda(t_k)=\int_{t_k}^{t_{k+1}}dt~\hat{a}^\dagger(-t)\hat{a}(-t)=\hat{n}_k$ is the photon number operator of the $k$-th time bin and $T$ denotes transpose. In practice, having the unitary operators in the exponential form is sufficient. 

The maximum bond dimension, defined as $\chi_{\rm max}=\max_{i_k}(D_{i_k})$, characterises the numerical difficulty of implementing these MPSs. Initially, the MPS description is exact up to the truncation of the input cavity Hilbert space, and the maximum bond dimension is $\chi_{\rm max}=1$. At total time evolution, and without any truncation of small singular values, the maximum bond dimension is $\chi_{\rm max}$ is equal to the Hilbert space dimension of the system (the cavities and TLS). In practice, we set a cutoff ($10^{-7}$) to truncate the small singular values to reduce the run-time cost. When truncating, one should set the orthogonality centre to either one of the adjacent matrices such that the singular values are the Schmidt coefficients of the state decomposed in the bipartite subspaces. We chose $\Gamma\Delta t =0.005$ and 2000 time bins.

We calculate two observables from the photon field: namely, the first-order coherence $G^{(1)}(t,t')=\expval{\hat{a}^\dagger(t)\hat{a}(t')}$ and projective measurements onto the single-photon sector. In the time bin representation, the field operators are mapped by $\hat{a}(t_k)\rightarrow \Delta B(t_k)/\Delta t$; hence, the first-order coherence $G^{(1)}(t_k,t_{k'})=\expval{\Delta B^\dagger(t_k)\Delta B(t_{k'})}/\Delta t^2$. The unnormalised density matrix after tracing over all single-photon detection times is expressed as $\hat{\rho}=\Tr_s{\sum_k\bra{1_k}\ket{\rm out}\bra{\rm out}\ket{1_k}}$, where $\ket{1_k}$ denotes the product state with $\ket{1}$ in the $k$th time bin and a vacuum in all other time bins, and $\Tr_s$ denotes tracing over the system. Ultimately, we numerically simulated the MPS description of the dynamics using the \textsc{Julia} package \texttt{ITensors.jl} \cite{itensor}.

\newpage

\section{Deterministic $\theta$ state generation}
\label{appendix: theta state generation}
\begin{figure}[!t]
    \centering
    \includegraphics[width=0.9\linewidth]{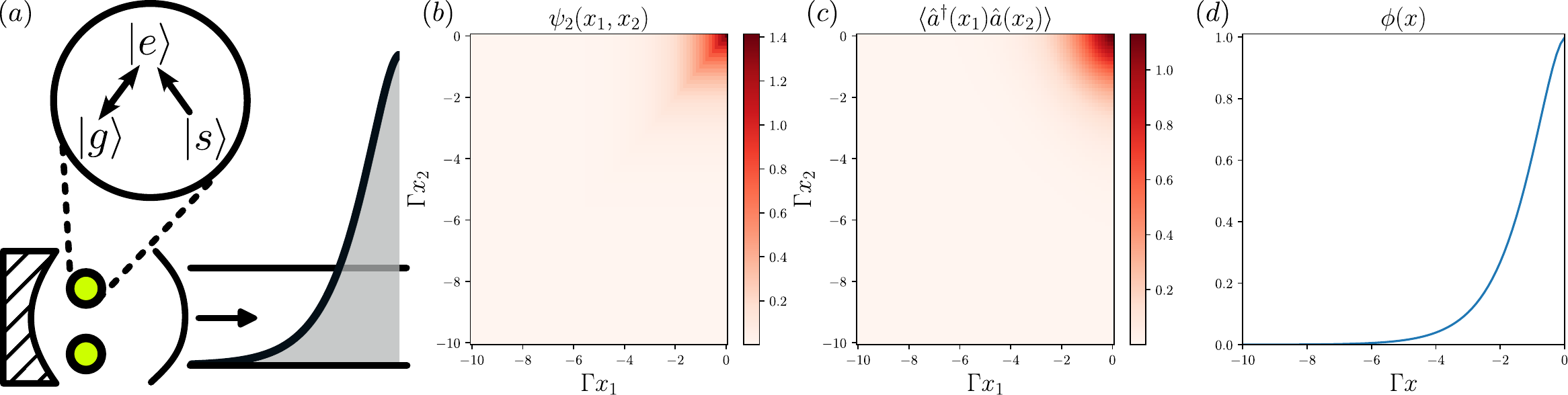}
    \caption{Deterministic $\theta$ state generation using two three-level atoms coupled to a cavity. (a) Three-level atoms each with ground state $|g \rangle$, excited state $|e\rangle$ and shelving state $|s \rangle$ coupled to a single-sided cavity. Photons emitted from the atoms couple to the cavity which is itself coupled to a collection mode. (b) Two-photon wavefunction $\psi(x_1, x_2)$ of a two-excitation superradiant state. (c) First-order correlation function $\langle \hat{a}^\dagger(x_1) \hat{a}(x_2)\rangle$ of the two-photon superradiant state. (d) Eigenvector $\phi(x)$ corresponding to the largest eigenvalue of $\langle \hat{a}^\dagger(x_1) \hat{a}(x_2)\rangle$.}
    \label{fig:deterministicSuperradiant}
\end{figure}

In this appendix, we provide further details on the deterministic generation of $\theta$ states using two three-level atoms coupled to a cavity. The cavity is then coupled to a one-dimensional continuum of modes. The three-level atoms have ground state $\ket{g}$, shelving state $\ket{s}$, and an optically excited state $\ket{e}$. This is shown in Fig.~\hyperref[fig:deterministicSuperradiant]{11(a)}. Let us take the ground and shelving states to be at zero energy. The Hamiltonian for the entire system is expressed as ($\hbar=v_g=1$)
\begin{align}
    \hat{H} &= \omega_0 \left[ \hat{\sigma}^1_+ \hat{\sigma}^1_- + \hat{\sigma}^2_+ \hat{\sigma}^2_- \right] + \omega_c \hat{c}^\dagger \hat{c} + \int dx \, \hat{a}^\dagger(x) \left[\tilde{\omega} - i \partial_x \right] \hat{a}(x)\\
    & - i \sqrt{\kappa} \left[\hat{a}^\dagger(0)\hat{c} - \hat{c}^\dagger \hat{a}(0) \right] + g \left[(\hat{\sigma}_+^1 + \hat{\sigma}_+^2) \hat{c} + \hat{c}^\dagger (\hat{\sigma}_-^1 + \hat{\sigma}_-^2) \right].
\end{align}
where $\sigma_-^i$ ($\sigma_+^i$) is the lowering (raising) operator for the $\ket{g}$-$\ket{e}$ transition of atom $i=1,2$, $\hat{c}$ ($\hat{c}^\dagger$) is the cavity annihilation (creation) operator and $\hat{a}(x)$ [$\hat{a}^\dagger(x)$] is the channel annihilation (creation) operator at position $x$. We also have $\kappa$ as the cavity-channel decay rate, $g$ is the atom-cavity coupling rate, $\omega_0$ is the atomic optical resonance frequency, $\omega_c$ is the cavity resonance frequency, and $\tilde \omega$ is the frequency about which we linearise the dispersion of the channel. 

Let us consider the atoms to be on resonance with the cavity and that the one-dimensional photon modes are linearised about this frequency, i.e., $\omega_0 = \omega_c=\tilde{\omega}$. Moving to an interaction picture with respect to the Hamiltonian $\hat{H}_0 = \omega_0 [\sigma_+^1 \sigma_-^1 + \sigma_+^2 \sigma_-^2 + \hat{c}^\dagger \hat{c} + \int dx \hat{a}^\dagger(x) \hat{a}(x)]$, we have the Hamiltonian in the interaction picture,
\begin{align}
    \hat{H}_I = -i \int dx \hat{a}^\dagger(x) \partial_x \hat{a}(x) - i \sqrt{\kappa} \left[ \hat{a}^\dagger (0) \hat{c} - \hat{c}^\dagger \hat{a}(0) \right] + g \left[ (\hat{\sigma}_+^1 + \hat{\sigma}_+^2) \hat{c} + \hat{c}^\dagger (\hat{\sigma}_-^1 + \hat{\sigma}_-^2)  \right].
\end{align}
The equations of motion for the photon channel and the cavity operators are
\begin{align}
    i \partial_t \hat{a}(x, t) &= - i \partial_x \hat{a}(x, t) - i\sqrt{\kappa} \delta(x) \hat{c}(t),\\
    i \partial_t \hat{c}(t) &= i \sqrt{\kappa} \hat{a}(0) + g (\hat{\sigma}_-^1 + \hat{\sigma}_-^2).
\end{align}
Solving for the channel operators, we have
\begin{equation}
    \hat{a}(x,t) = \hat{a}(x-t,0) - \sqrt{\kappa}\hat{c}(t - x)\theta(x)\theta(t-x).
\end{equation}
With this solution, the differential equation for the cavity operator becomes
\begin{equation}
   i \partial_t \hat{c}(t) = i \sqrt{\kappa} \left[\hat{a}_{\rm in}(t) - \frac{\sqrt{\kappa}}{2}\hat{c}(t)\right] + g (\hat{\sigma}_-^1 + \hat{\sigma}_-^2), 
\end{equation}
where we have defined $\hat{a}_{\rm in}(t) = \hat{a}(-t,0)$. We now consider the system in the bad-cavity limit where $\kappa \gg g$. In this limit the cavity can be adiabatically eliminated and $\partial_t \hat{c}(t) \sim 0$. This allows us to make the approximation
\begin{equation}
    \hat{c}(t) \sim \frac{\sqrt{\kappa} \hat{a}_{\rm in}(t) - i g (\hat{\sigma}_-^1 + \hat{\sigma}_-^2)}{\kappa/2}.
\end{equation}

Let us now consider a system operator $\hat{s}$, where $\hat{s}$ can be any operator for atom 1 and/or atom 2. Using the result from the adiabatic elimination of the cavity, we find $\hat{s}$ obeys the equations of motion
\begin{align}
    i \partial_t \hat{s}(t) &= g \left[ \left[\hat{s}, \hat{\sigma}_+^1 + \hat{\sigma}_+^2 \right] \hat{c} + \hat{c}^\dagger \left[\hat{s}, \hat{\sigma}_-^1 + \hat{\sigma}_-^2 \right] \right]\\
    i \partial_t \hat{s}(t) &\sim \frac{2g}{\sqrt{\kappa}} \left[ \left[\hat{s}, \hat{\sigma}_+^1 + \hat{\sigma}_+^2 \right] \left[ \hat{a}_{\rm in}(t) - i g (\hat{\sigma}_-^1 + \hat{\sigma}_-^2)\right] + \left[\hat{a}^\dagger_{\rm in}(t) + i g (\hat{\sigma}_+^1 + \hat{\sigma}_+^2) \right] \left[\hat{s}, \hat{\sigma}_-^1 + \hat{\sigma}_-^2 \right] \right].
\end{align}
This evolution equation can arise from an effective Hamiltonian:
\begin{equation}
    \hat{H}_{\rm eff} = -i \int dx \hat{a}^\dagger(x) \partial_x \hat{a}(x) + \sqrt{\Gamma} \left[\hat{a}^\dagger(0)(\hat{\sigma}_-^1 + \hat{\sigma}_-^2) + (\hat{\sigma}_+^1 + \hat{\sigma}_+^2) \hat{a}(0) \right].
\end{equation}
After adiabatic elimination of the cavity, we obtain an effective evolution under a Hamiltonian where the atom directly couples to the one-dimensional channel with effective decay rate $\Gamma = 4 g^2/\kappa$. 

To find the two-photon superradiant state we solve for the dynamics of the system under the effective Hamiltonian for the initial condition that the two atoms are initially in the excited state $\ket{ee}$. The superradiance problem is exactly solvable for $N$ atoms \cite{YudsonJETP1985}. Here we provide the two-atom two-excitation solution.  To do this we write the entire wavefunction of the system using a two-excitation ansatz:
\begin{equation}
    \ket{\psi(t)} = \left[ \frac{1}{\sqrt{2}}\int dx_1 dx_2 \hat{a}^\dagger(x_1) \hat{a}^\dagger(x_2) \alpha(x_1, x_2, t) + \int dx \hat{a}^\dagger(x) \hat{\sigma}_+^1 p_1(x,t) + \int dx \hat{a}^\dagger(x) \hat{\sigma}_+^2 p_2(x,t) + s(t) \hat{\sigma}_+^1 \hat{\sigma}_+^2 \right] \ket{0},
\end{equation}
where $\ket{0}$ indicates both atoms are in the ground state and the photon field in a vacuum. Note that the effective Hamiltonian is permutationally invariant under exchange of the atom indices $1 \leftrightarrow 2$. This means that $p_1(x,t) = p_2(x,t) \equiv p(x,t)$. From this ansatz we have the equations of motion
\begin{align}
    i \partial_t \alpha(x_1, x_2, t) &= - i \left[\partial_{x_1}  + \partial_{x_2} \right] \alpha(x_1, x_2, t) + \sqrt{2\Gamma} \left[\delta(x_1) p(x_2, t) + \delta(x_2) p(x_1, t) \right],\label{eq:alphaDyn}\\
    i \partial_t p(x,t) &= - i \partial_x p(x,t) + \sqrt{\Gamma} \delta(x) s(t) + \sqrt{\frac{\Gamma}{2}} \left[ \alpha(0,x,t) + \alpha(x,0,t) \right],\label{eq:pDyn}\\
    i \partial_t s(t) &= 2\sqrt{\Gamma} p(0,t),\label{eq:pDyn}
\end{align}
with the initial condition $p(x,0)=\alpha(x_1,x_2,0)=0$ and $s(0)=1$. By integrating (\ref{eq:alphaDyn}) and (\ref{eq:pDyn}) over the origin, we obtain  $p(0^+, t) = -i\sqrt{\Gamma}s(t)$ and $\alpha(0^+, x, t) = -i\sqrt{2\Gamma} p(x,t)$. To obtain these we used the definition $f(0) = [f(0^+) + f(0^-)]/2$ and the fact that the fields vanish for $x<0$. Using these, we can easily solve the third equation and obtain
\begin{align}
    s(t) = e^{-\Gamma t}.
\end{align}
To solve the equation for $p(x,t)$, we introduce comoving coordinates $\xi = x - t$ and $\eta=t$ and use the relation $\alpha(0^+, x, t) = -i\sqrt{2\Gamma} p(x,t)$. Formally integrating the equation, we obtain
\begin{equation}
    p(\xi, \eta) = -i \sqrt{\Gamma} e^{-\Gamma (\xi+\eta)} s(- \xi)\theta(-\xi)\theta(\eta + \xi) = -i \sqrt{\Gamma} e^{-\Gamma  \eta} \theta(-\xi)\theta(\eta + \xi).
\end{equation}
Finally, to solve for $\alpha(x_1, x_2, t)$ we again use comoving coordinates, $\xi_1 = x_1 - t$, $\xi_2 = x_2 - t$, $\eta =t$. After this transformation the equation can again be solved by integration, giving
\begin{align}
    \alpha(\xi_1, \xi_2, \eta) = - i \sqrt{2\Gamma} \left[ p(\xi_1, -\xi_2) \theta(-\xi_2)\theta(\xi_2+\eta) + p(\xi_2, -\xi_1)\theta(-\xi_1)\theta(\xi_1 +\eta) \right].
\end{align}
In the $t\rightarrow \infty$ limit, the atom has completely decayed and the entire photon wavefunction is away from $x=0$. Using comoving coordinates we define a two-photon wavefunction obtained from the two-excitation superradiant decay:
\begin{align}
    \ket{\bar{2}} = \frac{1}{\sqrt{2}} \int dx_1 dx_2 \hat{a}^\dagger(x_1) \hat{a}^\dagger(x_2) \ket{0} \psi(x_1, x_2),
\end{align}
with
\begin{equation}
    \psi(x_1, x_2) = \theta(-x_1) \theta(-x_2) \left[ e^{\Gamma x_2} \theta(x_1 - x_2) + e^{\Gamma x_1} \theta(x_2 - x_1)\right].
\end{equation}
The wavefunction is plotted in Fig.~\hyperref[fig:deterministicSuperradiant]{11(b)}. In the main text, the wavefunction is given in the $x_1 \geq x_2$ subspace.

\subsection{Calculation of the fidelity of the superradiant state}

The use of superradiant emission to make the $\theta$ state produces the state defined in the main text as
\begin{align}
    \ket{\bar{\theta}} = \cos{(\theta/2)} \ket{0} + \sin{(\theta/2)} \ket{\bar{2}}.
\end{align}
We wish to compute the fidelity of this state with the ideal $\theta$ state, i.e., $F = |\braket{\theta}{\bar{\theta}}|^2$.  Since the photon in the ideal $\theta$ state can be in any mode, we can calculate this overlap by decomposing $\ket{\bar{2}}$ into an orthogonal basis and choosing the dominant mode. There are two equivalent ways to do this: using the Takagi decomposition of the two-photon wavefunction $\psi(x_1, x_2)$ or diagonalising the first-order coherence function $\langle \hat{a}^\dagger(x_1) \hat{a}(x_2)\rangle$ of the two-photon wavefunction. We choose the latter. The first-order correlation function of the two-photon state is expressed as
\begin{equation}
    \langle \hat{a}^\dagger(x_1) \hat{a}(x_2)\rangle = \int dx \, \psi^*(x_1, x) \psi(x, x_2).
\end{equation}
This is plotted in Fig.~\hyperref[fig:deterministicSuperradiant]{11(c)}. For $x_1$ and $x_2$ as discrete variables, the first-order coherence function forms a Hermitian matrix whose eigenvectors give an orthonormal basis on which $\psi(x_1, x_2)$ can be represented. The eigenvector with the largest eigenvalues gives the dominant mode, which we denote $\phi(x)$, which is plotted in Fig.~\hyperref[fig:deterministicSuperradiant]{11(d)}. We then have
\begin{align}
    \braket{2}{\bar{2}} = \int dx_1 dx_2 \psi^*(x_1, x_2) \phi(x_1) \phi(x_2)
\end{align}
We find that for the superradiant state $\braket{2}{\bar{2}} = 0.954$. This then gives the overlap
\begin{align}
    F = |\braket{\theta}{\bar \theta}|^2 = |\cos^2{(\theta/2)} + \sin^2{(\theta/2)} \braket{2}{\bar 2}|^2 &= 1 - 2 \sin^2{(\theta/2)}\operatorname{Re}{\left[ 1 - \braket{2}{\bar 2}\right]} + 4 \sin^4{\theta/2} |1-\braket{2}{\bar 2}|^2\\
    &\sim 1 - 2 \sin^2{(\theta/2)} \times 0.046. 
\end{align}

\subsection{Inclusion of loss in superradiant decay}

Here we calculate the effect of imperfect coupling of the atoms to the one-dimensional channel on the fidelity of the $\theta$ states. We do this by considering a worst-case scenario where the loss is a fraction of the superradiant decay. To do this we add an additional channel of modes with coupling coefficient $\gamma$ in the effective Hamiltonian:
\begin{align}
    \hat{H}_{\rm eff} &= -i \int dx \hat{a}^\dagger(x) \partial_x \hat{a}(x)  -i \int dx \hat{b}^\dagger(x) \partial_x \hat{b}(x) + \sqrt{\Gamma} \left[\hat{a}^\dagger(0)(\hat{\sigma}_-^1 + \hat{\sigma}_-^2) + (\hat{\sigma}_+^1 + \hat{\sigma}_+^2) \hat{a}(0) \right]\\
    &+ \sqrt{\gamma} \left[\hat{b}^\dagger(0)(\hat{\sigma}_-^1 + \hat{\sigma}_-^2) + (\hat{\sigma}_+^1 + \hat{\sigma}_+^2) \hat{b}(0) \right].
\end{align}
We now define new operators,
\begin{align}
    \hat{a}_{e}(x) &= \frac{\sqrt{\Gamma} \hat{a}(x) + \sqrt{\gamma}\hat{b}(x)}{\sqrt{\Gamma + \gamma}} = \sqrt{\beta}\hat{a}(x) + \sqrt{1 - \beta} \hat{b}(x),\\
    \hat{a}_{o}(x) &= \frac{\sqrt{\gamma} \hat{a}(x) - \sqrt{\Gamma}\hat{b}(x)}{\sqrt{\Gamma + \gamma}} = \sqrt{1-\beta}\hat{a}(x) - \sqrt{\beta} \hat{b}(x),\\
\end{align}
where we have defined the coupling efficiency $\beta = \Gamma/(\Gamma + \gamma)$. With these operators the effective Hamiltonian becomes
\begin{align}
    \hat{H}_{\rm eff} &= -i \int dx \hat{a}_e^\dagger(x) \partial_x \hat{a}_e(x)  -i \int dx \hat{a}_o^\dagger(x) \partial_x \hat{a}_o(x) + \sqrt{\Gamma+\gamma} \left[\hat{a}_e^\dagger(0)(\hat{\sigma}_-^1 + \hat{\sigma}_-^2) + (\hat{\sigma}_+^1 + \hat{\sigma}_+^2) \hat{a}_e(0) \right].
\end{align}
We note that the $\hat{a}_o(x)$ operator does not enter as an interaction term and is decoupled from the system. The effective Hamiltonian is therefore identical to the original one, but with $\Gamma \rightarrow \Gamma + \gamma$ and $\hat{a} \rightarrow \hat{a}_e$. The two-excitation superradiant solution is thus,
\begin{align}
    \ket{\bar{2}}_e &= \frac{1}{\sqrt{2}} \int dx_1 dx_2 \hat{a}_e^\dagger(x_1) \hat{a}_e^\dagger(x_2) \ket{0} \psi(x_1, x_2)\\
    &= \frac{1}{\sqrt{2}} \int dx_1 dx_2 \left[\sqrt{\beta}\hat{a}^\dagger(x_1) + \sqrt{1-\beta}\hat{b}^\dagger(x_1)\right]\left[\sqrt{\beta}\hat{a}^\dagger(x_2) + \sqrt{1-\beta}\hat{b}^\dagger(x_2)\right] \ket{0} \psi(x_1, x_2),
\end{align}
with
\begin{equation}
    \psi(x_1, x_2) = \theta(-x_1) \theta(-x_2) \left[ e^{(\Gamma+\gamma) x_2} \theta(x_1 - x_2) + e^{(\Gamma+\gamma) x_1} \theta(x_2 - x_1)\right].
\end{equation}
The wavefunction therefore has the same shape, but with the effective decay rate scaling with the sum of the decay of the two channels $\Gamma + \gamma$. We can now compute the fidelity as before,
\begin{align}
    F = |\braket{\theta}{\bar \theta}|^2 = |\cos^2{(\theta/2)} + \sin^2{(\theta/2)} \braket{2}{\bar 2}_e|^2 &= 1 - 2 \sin^2{(\theta/2)}\operatorname{Re}{\left[ 1 - \braket{2}{\bar 2}_e\right]} + 4 \sin^4{\theta/2} |1-\braket{2}{\bar 2}_e|^2\\
    &\sim 1 - 2 \sin^2{(\theta/2)} (1-0.954\beta),
\end{align}
where we assume that the loss is small such that  $(1-\beta) \ll 1$.

\end{document}